\documentclass[11pt, oneside, a4paper]{article} 
\usepackage{fourier} 
\usepackage[english]{babel} 
\usepackage{lineno}
\usepackage{authblk}

\usepackage{amsmath, tikz, caption, subcaption, float,microtype,commath}

\usetikzlibrary{matrix,arrows,decorations.pathmorphing}
\usetikzlibrary{positioning}
\usepackage[top=2.54cm, bottom=2.54cm, left=3.17cm, right=3.17cm, showframe=false]{geometry}

\DisableLigatures{encoding = *, family = *}

\usepackage{fancyhdr} % Required for custom headers
\numberwithin{equation}{section}

\title{Bacterial fitness shapes the population dynamics of antibiotic-resistant and -susceptible bacteria in a model of combined antibiotic and anti-virulence treatment}

\author[a,*]{Lucy Ternent}

\author[a]{Rosemary J. Dyson}

\author[b]{Anne-Marie Krachler} 

\author[a,b,**]{Sara Jabbari}

\affil[a]{School of Mathematics, University of Birmingham, United Kingdom}

\affil[b]{Institute of Microbiology and Infection, University of Birmingham, United Kingdom}

\affil[*]{Present address: Molecular Organisation and Assembly in Cells Doctoral Training Centre, University of Warwick, United Kingdom}

\affil[**]{Address correspondence to Sara Jabbari, s.jabbari@bham.ac.uk, +44 121 4146196}

\date{Submitted to Journal of Theoretical Biology, July 21st 2014}

\begin{document}

%\linenumbers
\maketitle
\begin{abstract}

\noindent Bacterial resistance to antibiotic treatment is a huge concern: introduction of any new antibiotic is shortly followed by the emergence of resistant bacterial isolates in the clinic. 
This issue is compounded by a severe lack of new antibiotics reaching the market.
The significant rise in clinical resistance to antibiotics is especially problematic in nosocomial infections, where already vulnerable patients may fail to respond to treatment, causing even greater health concern. 
A recent focus has been on the development of anti-virulence drugs as a second line of defence in the treatment of antibiotic-resistant infections.
This treatment, which weakens bacteria by reducing their virulence rather than killing them, %rather than killing the bacteria, would weaken them by reducing their virulence, 
should allow infections to be cleared through the body's natural defence mechanisms. In this way there should be little to no selective pressure exerted on the organism and, as such, a predominantly resistant population would be unlikely to emerge. 
However, much controversy surrounds this approach with many believing it would not be powerful enough to clear existing infections, restricting its potential application to prophylaxis.
We have developed a mathematical model that provides a theoretical framework to reveal the circumstances under which anti-virulence drugs may or may not be successful. We demonstrate that by harnessing and combining the advantages of antibiotics with those provided by anti-virulence drugs, given infection-specific parameters, it is possible to identify treatment strategies that would efficiently clear bacterial infections, while preventing the emergence of resistant subpopulations. Our findings strongly support the continuation of research into anti-virulence drugs and demonstrate that their applicability may reach beyond infection prevention.
\end{abstract}


\section{Introduction}

Bacterial resistance to antibiotic agents is an increasing problem in modern society. 
The introduction of every new class of antibiotic, from the original $\beta$-lactam, penicillin, to the more recent lipopeptides such as daptomycin, has been followed by the emergence of new strains of bacteria resistant to that class, many emerging in the clinic only a few years after the introduction of the drug \cite{Butler:2006aa, Clatworthy:2007aa, Lewis:2013aa}. 
Given that the pace of antibiotic discovery has dramatically slowed down (most classes of antibiotic were discovered in the 1940s to the 1960s, the `Golden Era' of antibiotics, with the past forty years giving us only five significant new classes \cite{Butler:2006aa, Lewis:2013aa} and pharmaceutical companies devoting less research into discovering new antibiotics \cite{Mellbye:2011aa, Projan:2004aa}), this poses a huge problem. 
Thus, development of novel treatments for bacterial infections is of utmost importance.

Traditional antibiotics are either classed as bacteriocidal or bacteriostatic, working to either kill bacteria or inhibit bacterial growth respectively \cite{Clatworthy:2007aa, Mellbye:2011aa}. 
While effective in eliminating susceptible bacterial infections, antimicrobials impose selective pressure on the bacteria, leading to the rise of resistant clones within the bacterial population. 
Resistance can be acquired either through spontaneous chromosomal mutation and then selection by antibiotic use, known as vertical evolution, or through acquiring genetic material from other resistant organisms, known as horizontal evolution \cite{Tenover:2006aa}. 
Horizontal evolution occurs via mechanisms such as conjugation, transformation and transduction \cite{Alanis:2005aa} and can take place between strains of the same or different bacterial species.
Horizontal resistance can lead to multidrug resistance and is a major concern in hospitals, where resistant bacteria are able to remain viable despite antibiotic use, and are the cause of many serious nosocomial infections in already vulnerable patients \cite{Alanis:2005aa, Tenover:2006aa}.

It has been suggested that the focus of new drug development should be on targeting virulence, the bacteria's ability to cause disease \cite{Clatworthy:2007aa}, but this approach remains controversial. 
Anti-virulence drugs would minimize any harm caused by bacteria while they remain in the host until they can be cleared by natural defences.  This can occur either by being flushed out of the system or by being eradicated by the host's immune response. 
This \textit{should} exert little to no selective pressure on the bacteria but this has not yet been proved \textit{in vivo}. 
Anti-virulence drugs could target a range of mechanisms, including bacterial adhesion to host cells, toxin delivery or virulence gene regulation, all necessary for successful infection \cite{Clatworthy:2007aa, Mellbye:2011aa}.

We adopt a modelling approach to investigate the viability of anti-virulence drugs \textit{in silico}.
Our results suggest that, in combination with antibiotics and under specific treatment strategies, anti-virulence drugs can be effective in treating bacterial infections where antibiotic resistance is a concern.
Optimal treatment strategies are likely to be specific to the patient, infection and bacterial strain and (rather than attempt to pin-point exact strategies) we use this theoretical framework as a ``proof of concept'', exploiting parameter surveys to investigate a range of scenarios, highlighting the potential (and possibly the need) for addressing infections with tailored treatments in the future.
Given that every patient, infection and strain of bacteria are different, it is impossible to obtain a conclusive set of parameters which will suit all situations.
We therefore exploit parameter surveys to ascertain under what conditions certain behaviour will occur.

\section{Model formulation} \label{MathModelResistance}

%\paragraph{}

There are two main approaches usually taken when modelling the emergence of antibiotic resistant bacteria: within-host models, or within-hospital compartmental models. 
Models of hospital resistance usually follow a similar form to the classic ``Susceptible-Infectious-Resistant'' (SIR) models of epidemiology, looking specifically at how nosocomial infections will spread throughout the hospital, for example \cite{Austin:1999aa, Lipsitch:2000aa, Webb:2005aa}. 
While such models are useful to develop strategies to prevent the spread of resistance, our focus is on treatment strategies to eliminate the emergence of a resistant subpopulation that has either arisen through random mutation and clonal expansion or through cross-contamination within a particular infection and under a specific treatment regimen. If the resistant bacteria within a single host can be eliminated then the spread of resistance throughout the hospital is a less pressing concern.

Existing mathematical models that focus on the within-host emergence of resistance examine how antibiotic treatment strategies can both cause and be adapted to prevent the emergence of antibiotic resistance, for example \cite{DAgata:2008aa,Lipsitch:1997aa}.
Such models often neglect the effect of the host immune response, assuming it to be negligible or constant under the effect of the antibiotic.
Since the efficacy of anti-virulence drugs will depend at least in part on the host's innate immunity, we include cell-mediated innate immune response in the model.

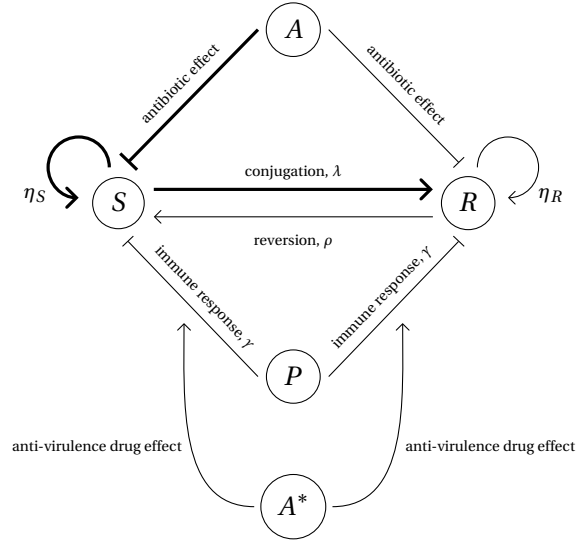
\begin{figure}[ht]
\begin{center}
\begin{tikzpicture}[>=angle 90,descr/.style={fill=white},text height=1.5ex, text depth=0.25ex,scale=1.15]
\node (a) at (2.2,4.5) {};
\node (s) at (0.5,3) {};
\node (-a) at (2.8,4.5) {};
\node (-s) at (0.5,2.1) {};
\node (r) at (4.5,3) {};
\node (-r) at (4.5,2.1) {};
\node (i) at (2.2,0.5) {};
\node (-i) at (2.8,0.5) {};
\node (+s) at (0.8,2.5) {};
\node (+r) at (4.2,2.5) {};
\node (r1) at (4.8,2.5) {};
\node (etaS) at (-0.45,2.65) {{\scriptsize $\eta_S$}};
\node (etaR) at (5.45,2.65) {{\scriptsize $\eta_R$}};
\node (a1) at (1.25,1.35) {};
\node (a2) at (3.75,1.35) {};
\path (0.5,2.5) node[draw,shape=circle] (S) {$S$};
\path (4.5,2.5) node[draw,shape=circle] (R) {$R$};
\path (2.5,0.5) node[draw,shape=circle] (I) {$P$};
\path (2.5,4.5) node[draw,shape=circle] (A) {$A$};
\path (2.5,-1) node[draw,shape=circle] (Astar) {$A^*$};
\path [->,font=\tiny]
([xshift=-2pt]Astar.west) edge [out=180, in=270]  node [midway, left] {anti-virulence drug effect} (a1);
\path [->, font=\tiny]
([xshift=2pt]Astar.east) edge [out=0, in=270] node [midway, right] {anti-virulence drug effect} (a2);
\path [-|, very thick,font=\tiny]
(a.west) edge node [midway, sloped, above] {antibiotic effect} ([yshift= -1pt]s.east) ; 
\path [-|,font=\tiny]
(-a.east) edge node [midway, sloped, above] {antibiotic effect}  ([yshift= -1pt]r.west);
\path [-|,font=\tiny]
(i.west) edge node [midway, sloped, above] {immune response, $\gamma$} ([yshift= -1pt]-s.east);
\path [-|,font=\tiny]
(-i.east) edge node [midway, sloped, above] {immune response, $\gamma$} ([yshift= -1pt]-r.west);
\path [->, very thick,font=\tiny]
([yshift= 4.5pt]+s.east) edge  node [midway, sloped, above] {conjugation, $\lambda$} ([yshift= 4.5pt]+r.west);
\path [<-,font=\tiny]
([yshift= -4.5pt]+s.east) edge  node [midway, sloped, below] {reversion, $\rho$} ([yshift= -4.5pt]+r.west);
\draw [->,font=\tiny, very thick]
([yshift =-2.5pt]s.west) arc [start angle=0, end angle=270, radius=.35];
\draw [->, xscale=-1,font=\tiny]
([yshift =-2.5pt]r.east) arc [start angle=0, end angle=270, radius=.35];
\end{tikzpicture}
\end{center}
\caption{Schematic representation of the main interactions involved in an infection treated by antibiotics and anti-virulence drugs with $S$ (antibiotic-susceptible bacteria), $R$ (antibiotic-resistant bacteria), $P$ (immune cells, e.g. phagocytes), $A$ (antibiotic concentration) and $A^*$ (anti-virulence drug concentration). 
Both the phagocytes and antibiotic inhibit bacterial growth, with the antibiotic having a greater effect on the susceptible bacteria than the resistant ones, assuming only partial resistance.
The anti-virulence drug increases the effectiveness of the immune response in order to eliminate the bacteria more naturally, working on both susceptible and resistant bacteria in the same way.
} \label{ResistanceSchematic}
\end{figure}

The system consists of five interacting components: populations of antibiotic-susceptible bacteria ($S$), antibiotic-resistant bacteria ($R$) and immune cells e.g. phagocytes ($P$), and concentrations of antibiotic ($A$) and anti-virulence drug ($A^*$).  These interact as demonstrated in Figure \ref{ResistanceSchematic} and represent a generalised model of a local bacterial infection, such as a urinary tract or wound infection. 

%Five variables are required:
%\begin{itemize}
%\item{antibiotic-susceptible bacteria, $S$}
%\item{antibiotic-resistant bacteria, $R$}
%\item{antibiotic, $A$}
%\item{anti-virulence drug, $A^*$}
%\item{antibiotic-resistant bacteria, $R$}
%\end{itemize}
%These interact as demonstrated in Figure \ref{ResistanceSchematic} and represent a generalised model of a local bacterial infection, such as a urinary tract or wound infection. 

The growth of bacteria within a host is likely to saturate over time, as a result of space and nutrient limitations.  We therefore use a logistic growth term with baseline growth rate $\eta_i$, {\footnotesize$(i = S,R)$} and a combined carrying capacity $K$, rather than simple exponential growth, to represent the bacterial dynamics. 
We assume natural death of the bacterial cells is negligible relative to the effects of the drugs and the immune system, but we do include a removal rate, $\psi$, representing the body's endogenous, physical clearance mechanisms (this can be expected to vary depending upon infection type).

Due to the potential for multidrug resistance \cite{Tenover:2006aa}, the primary cause for concern in hospitals is resistance due to horizontal evolution: acquiring new genetic material from other resistant organisms. Horizontal evolution involves the transfer of the resistance gene, normally found in sections of DNA known as transposons, from one plasmid to another.  This takes place via one of three main mechanisms: conjugation, transformation or transduction \cite{Tenover:2006aa, Alanis:2005aa}. 
We focus on the most common of the three transmission mechanisms \cite{Alanis:2005aa}, conjugation, whereby one bacterium transfers plasmid containing the genes for resistance to an adjacent bacterium \cite{Tenover:2006aa}. 
Research suggests that these plasmid-bearing, and thus antibiotic-resistant, bacteria are subject to a fitness cost, $c$, lowering their growth rate  \cite{Austin:1999aa, Lipsitch:2000aa, Bergstrom:2000aa, Levin:1997aa, Lipsitch:2001aa}, hence we choose $\eta_R = \left( 1 - c \right) \eta_S$ where $0<c<1$. 
Since this plasmid transfer occurs between adjacent bacteria, and we assume a well mixed population, we represent this interaction through mass action kinetics with a conjugation rate, $\lambda$, proportional to the levels of both antibiotic-susceptible and -resistant bacteria in the population \cite{Webb:2005aa, DAgata:2008aa, Bergstrom:2000aa, Imran:2006aa}. 
It has also been observed that it is possible to lose the plasmid carrying the resistance genes \cite{Webb:2005aa, Imran:2006aa} and so this too is incorporated into the model via a constant reversion rate, $\rho$.

We assume that bacteria are consumed by phagocytes (immune cells) at rate $\gamma$. 
Rather than assuming a constant phagocyte level, as is often done, we incorporate a more realistic representation of cell-based, innate immune response into the model.
Using a logistic style term, phagocytes are recruited to the site of infection at a rate proportional to the amount of bacteria present subject to an infection-site-specific, phagocytic carrying capacity, $P_{\text{max}}$ \cite{Smith:2011aa}. 
Phagocytes are lost through pathogen-induced apoptosis (at rate $\delta$) and natural clearance (at rate $\delta_P$)  \cite{Zysk:2001aa}.

Antibiotic is either administered in one dose or assumed to be present at constant level (e.g. intravenous delivery) and rapidly absorbed to the site of infection \cite{Austin:1999aa}. 
Elimination of the antibiotic from the system, due to both degradation of the drug and clearance due to natural flow through the body, is assumed to be independent of the bacterial population and follows first order kinetics \cite{Austin:1999aa, Imran:2006aa}. The effect of the antibiotic on each bacteria type is modelled using an $E_{\text{max}}$ saturating response, $\frac{E_{\text{max}}^iA}{{A_{50}}^i + A}$, subject to a maximum killing rate $E_{\text{max}}^{i}$ and the antibiotic concentration required for half maximum effect, $A_{50}^{i}$, for $i = S, \, R$,\cite{Austin:1999aa, Lipsitch:1997aa, Imran:2007aa,  Nikolaou:2006aa}. 
Note that resistant bacteria are in reality often only partially resistant, thus in general we assume $E_{\text{max}}^{R}\neq{0}$ but $E_{\text{max}}^{R}<E_{\text{max}}^{S}$ \cite{Lipsitch:1997aa}.

We assume the anti-virulence drug increases the effectiveness of the host's immune response by weakening the bacteria's ability to counteract the host's immune mechanisms. For example, many bacterial pathogens use type III secreted effector proteins to reprogram cellular pathways including those associated with host defence \cite{Navarro:2005, Krachler:2011}.
In order to model this mathematically we base the effect of the anti-virulence drug on a saturating response, or Hill equation \cite{Csajka:2006aa, Goutelle:2008aa}, with $\gamma_{\text{max}}$ representing the maximum increased effect of the immune response and $\gamma_{50}$, the anti-virulence drug concentration for half maximum effect. 
We assume that the anti-virulence drug has the same effect on both the antibiotic-resistant and susceptible bacteria.

Combining the above, we obtain the following model of bacterial population dynamics during antibiotic treatment:

\begin{align}
&\dod{A}{t} \quad= \quad-\quad \alpha A, \label{Model1Antibiotic}\\
&\dod{A^*}{t}  =  \quad  - \quad \kappa A^*, \label{antimicrobial}\\
&\dod{P}{t} \quad = \quad \beta\left(S + R\right)\left( 1 - \frac{P}{P_{\text{max}}}\right) \quad- \quad \delta(S + R)P\quad - \quad \delta_P P,\\
&\dod{S}{t}\quad=\quad \eta_S S \left( 1 - \frac{S+R}{K}\right) \quad - \quad \mu_S (A)S \quad - \quad\left( \gamma + \zeta \left(A^*\right) \right) P S \quad- \quad \lambda SR \quad+ \quad \rho R \quad - \quad \psi S,\\
&\dod{R}{t} \quad  = \quad (1-c)\eta_S R \left(1-\frac{S+R}{K}\right) \quad- \quad\mu_R(A)R \quad- \quad \left( \gamma + \zeta \left(A^*\right) \right) P R \quad+ \quad \lambda SR\quad- \quad\rho R\quad  -\quad \psi R.\label{Req}
\end{align}
Here $\mu_i(A) = \dfrac{ E^{i}_{\text{max}}A}{A^i_{50} + A}$ is the effect of the antibiotic on susceptible and resistant bacteria, respectively, for $i = S, \, R$, and $\zeta \left( A^* \right) = \dfrac{{\gamma_{\text{max}} A^*}}{{\gamma_{50}} + {A^*}}$ is the effect of the anti-virulence drug.
Default parameter values and definitions are as given in Table \ref{Table1}, but a range of parameter values are explored throughout. 
Unless otherwise specified, we use the initial conditions:

\begin{align}
A(0)&=4 \hspace*{0.1cm} \mu\textrm{g/ml}, & A^*(0) &= 4\hspace*{0.1cm} \mu\textrm{g/ml}, & P(0) &= 0\hspace*{0.1cm} \textrm{cells}, & S(0) &= 6000\hspace*{0.1cm} \textrm{cells}, & R(0) &= 20\hspace*{0.1cm} \textrm{cells},
\end{align} 
representing clinically realistic dosages (for instance 4 $\mu$g/ml is a value for meropenem in \textit{Pseudomonas aeruginosa} infections \cite{LMD02}) and the antibiotic-resistant subpopulation being in the minority and introduced via cross-contamination.
We note that the initial conditions for numbers of bacteria are in line with the two studies from which we have principally sourced our parameter values \-- see Table \ref{Table1} and \cite{Smith:2011aa, Handel:2009aa}; given that we are intending to provide a framework for future development of infection-specific treatment strategies and not positing exact strategies here, this is sufficient for our purposes.
The equations are solved numerically in Matlab using the solver \textit{ode15s}.
We consider three scenarios in our numerical solutions: an infection is treated by antibiotics alone, by an anti-virulence drug alone or by both drugs in combination.
In each case we wish to see how effective the treatment strategies are in lowering bacterial load but crucially also in tackling the emergence of antibiotic-resistant bacteria during the infection.
 
\begin{table}[H]
{ \scriptsize
\caption{Parameter notation with descriptions, estimated values and units.
Immune-related parameter values are taken from \cite{Smith:2011aa} and bacteria and antibiotic ones from \cite{Handel:2009aa}.
Where no source is given estimates have been made from what we believe to be realistic parameter ranges, in keeping with the other parameters. In particular $\beta$ is chosen so that the immune system alone cannot clear the infection, the anti-virulence drug is assumed to have equivalent efficiency to the antibiotic, and $\lambda$ and $\rho$ are chosen so that an antibiotic-resistant strain can emerge to become dominant during infection. 
While we do not claim that these parameters represent one particular infection-type, they should fall within a realistic range, enabling us to investigate a spread of scenarios through variations to these values in the parameter surveys provided throughout the study.
\label{Table1}}
\begin{tabular}{l l l l l}
\hline %\\ [-1.75ex]
\textbf{Parameter} & \textbf{Description} & \textbf{Units} & \textbf{Estimated Value} & \textbf{Source}\\ [0.5ex] \hline \hline \\ [-1.75ex]
$\alpha$ & Elimination rate of antibiotic under distinct doses & days$^{-1}$ & 3.6 &  \cite{Handel:2009aa} \\ [0.5ex]
 & Elimination rate of antibiotic under intravenous therapy & days$^{-1}$ & 0 &  - \\ [0.5ex]
$\kappa$ & Elimination rate of anti-virulence drug & days$^{-1}$ & 3.6 &  - \\ [0.5ex]
 & Elimination rate of anti-virulence drug under intravenous therapy & days$^{-1}$ & 0 &  - \\ [0.5ex]
$\beta$ &Recruitment rate of phagocytes & days$^{-1}$ & $ 3 $ & - \\ [0.5ex]
$c$ & Fitness cost of resistance & dimensionless & 0.1 & - \\ [0.5ex]
$\delta_{P}$ & Clearance rate of phagocytes & days$^{-1}$ & $1.512$ & \cite{Smith:2011aa} \\ [0.5ex]
$\delta$ & Bacterial induced death of phagocytes & cells$^{-1}$days$^{-1}$ & $6 \times 10^{-6}$ & \cite{Smith:2011aa} \\ [0.5ex]
$A^{S}_{50}$ &Antibiotic concentration for half maximum effect on susceptible bacteria& $\mu$g/ml & 0.25 & \cite{Handel:2009aa} \\ [0.5ex]
$A^{R}_{50}$ &Antibiotic concentration for half maximum effect on resistant bacteria& $\mu$g/ml & 5 & \cite{Handel:2009aa}\\ [0.5ex]
$E^{S}_{\text{max}}$ & Maximum killing rate of susceptible bacteria &days$^{-1}$  & 36 &  \cite{Handel:2009aa}  \\ [0.5ex]
$E^{R}_{\text{max}}$ & Maximum killing rate of resistant bacteria &days$^{-1}$  & 26.4 &  \cite{Handel:2009aa}  \\ [0.5ex]
$\gamma_{50}$ &Anti-virulence drug concentration for half maximum effect & $\mu$g/ml & 5 & - \\ [0.5ex]
$\gamma_{\text{max}}$ & Maximum increased effectiveness of immune response &cells$^{-1}$ days$^{-1}$  & 0.035 &  -  \\ [0.5ex]
$h$ & Hill coefficient & dimensionless& $1$ & - \\ [0.5ex]
$K$ & Carrying capacity of bacteria & cells  & 10$^{9}$ & \cite{Handel:2009aa} \\ [0.5ex]
$\eta_S$ & Growth rate of susceptible bacteria & days$^{-1}$  & $24$ & \cite{Handel:2009aa} \\ [0.5ex]
$P_{\text{max}}$ & Maximum number of phagocytes & cells  & $1.8 \times 10^{5}$ & \cite{Smith:2011aa} \\ [0.5ex]
$\gamma$ & Bacterial clearance by phagocytes & cells$^{-1}$days$^{-1}$  & $2.4 \times 10^{-4}$ & \cite{Smith:2011aa} \\ [0.5ex]
$\lambda$ &Conjugation rate constant &days$^{-1}$ & $10^{-4}$ &  - \\ [0.5ex]
$\rho$ & Reversion rate constant & days$^{-1}$  & $10^{-6}$ &  - \\ [0.5ex]
$\psi$ & Removal rate & days$^{-1}$  & $0.7$ &  - \\ [0.5ex]
\hline
\end{tabular}
}
\end{table}

\section{Results}

\paragraph{Antibiotic treatment ($\mathbf{A(0)=4}$ $\mathbf{\mu}$g/ml, $\mathbf{A^*(0)=0}$ $\mathbf{\mu}$g/ml)\\}

\begin{figure}
\centering
\begin{subfigure}[b]{0.485\textwidth}
\includegraphics[width=\textwidth]{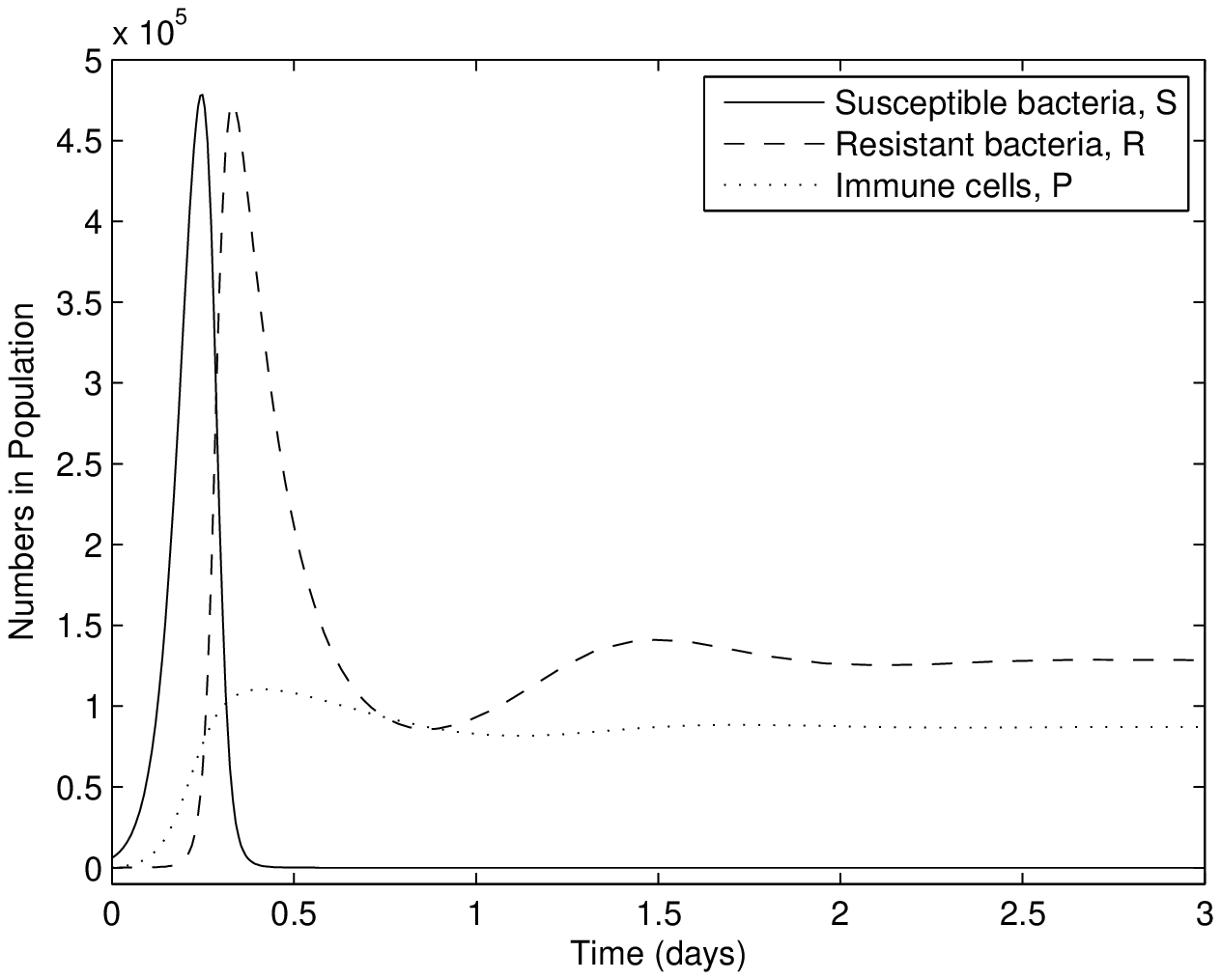}
\caption{}
\label{InitialSimulationNoAntibiotic}
\end{subfigure}
\quad
\begin{subfigure}[b]{0.485\textwidth}
\includegraphics[width=\textwidth]{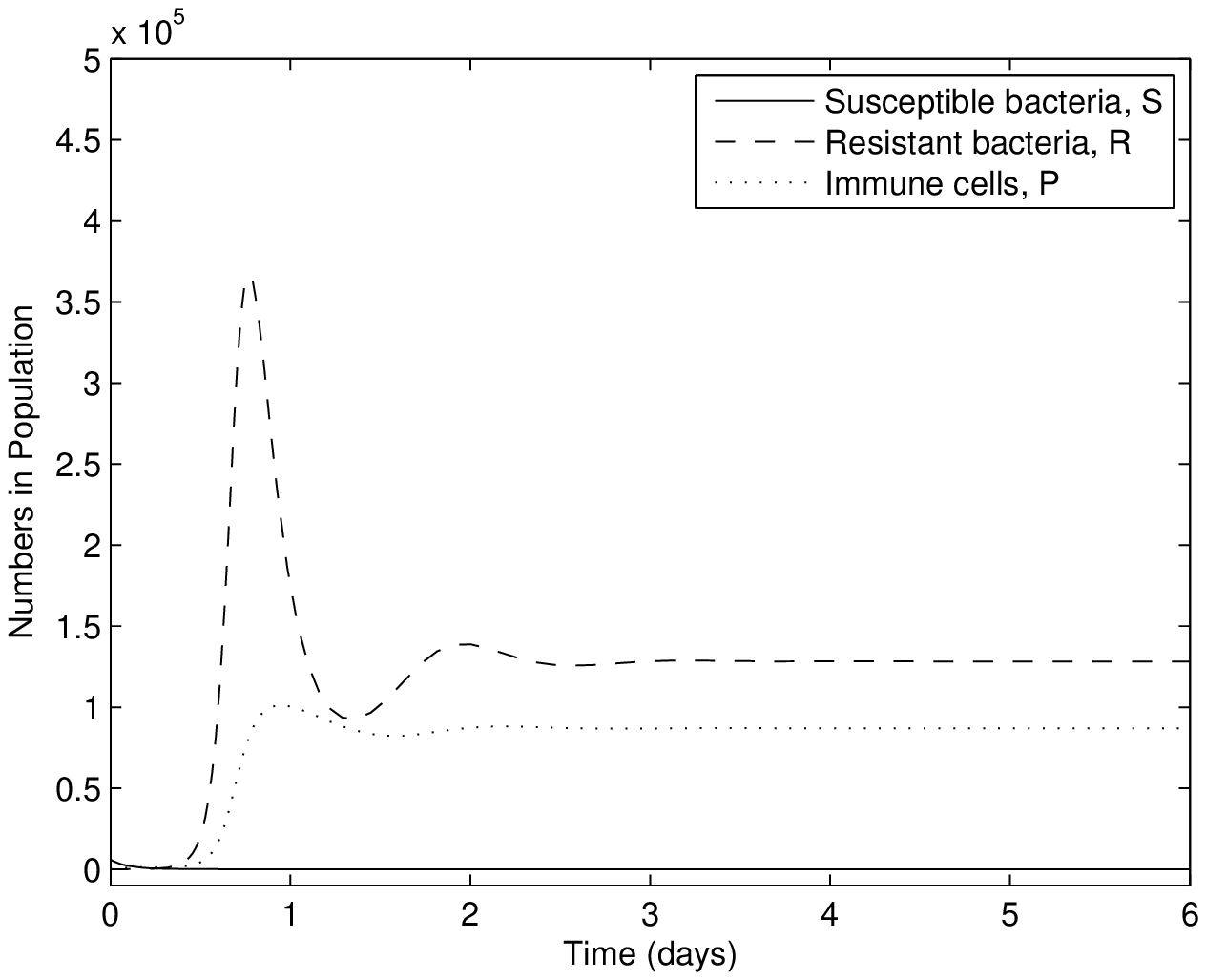}
\caption{}
\label{DegradingAntibioticModel1}
\end{subfigure}
\quad
\begin{subfigure}[b]{0.485\textwidth}
\includegraphics[width=\textwidth]{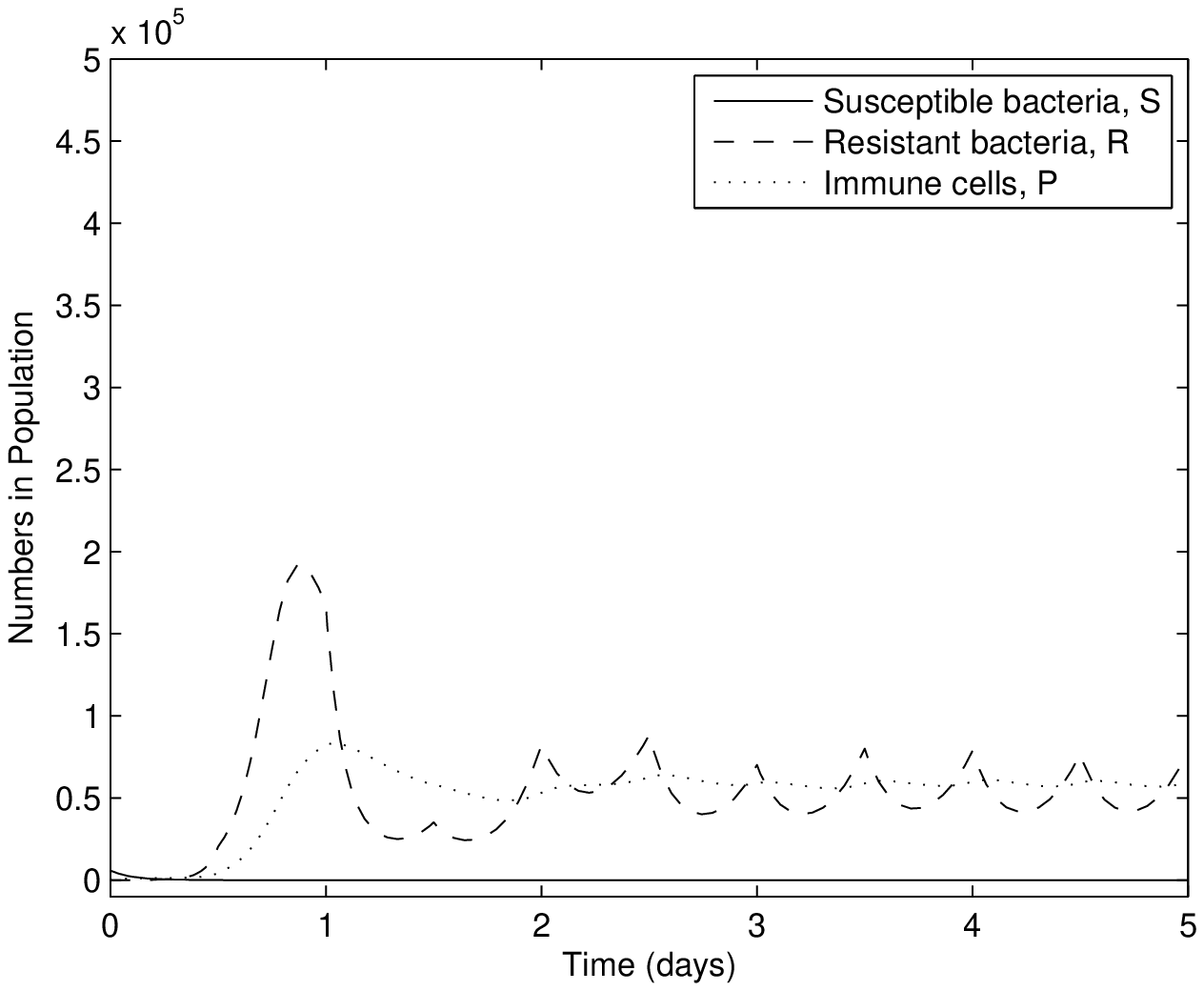}
\caption{}
\label{Model1DosingSimulation}
\end{subfigure}
\quad
\begin{subfigure}[b]{0.485\textwidth}
\includegraphics[width=\textwidth]{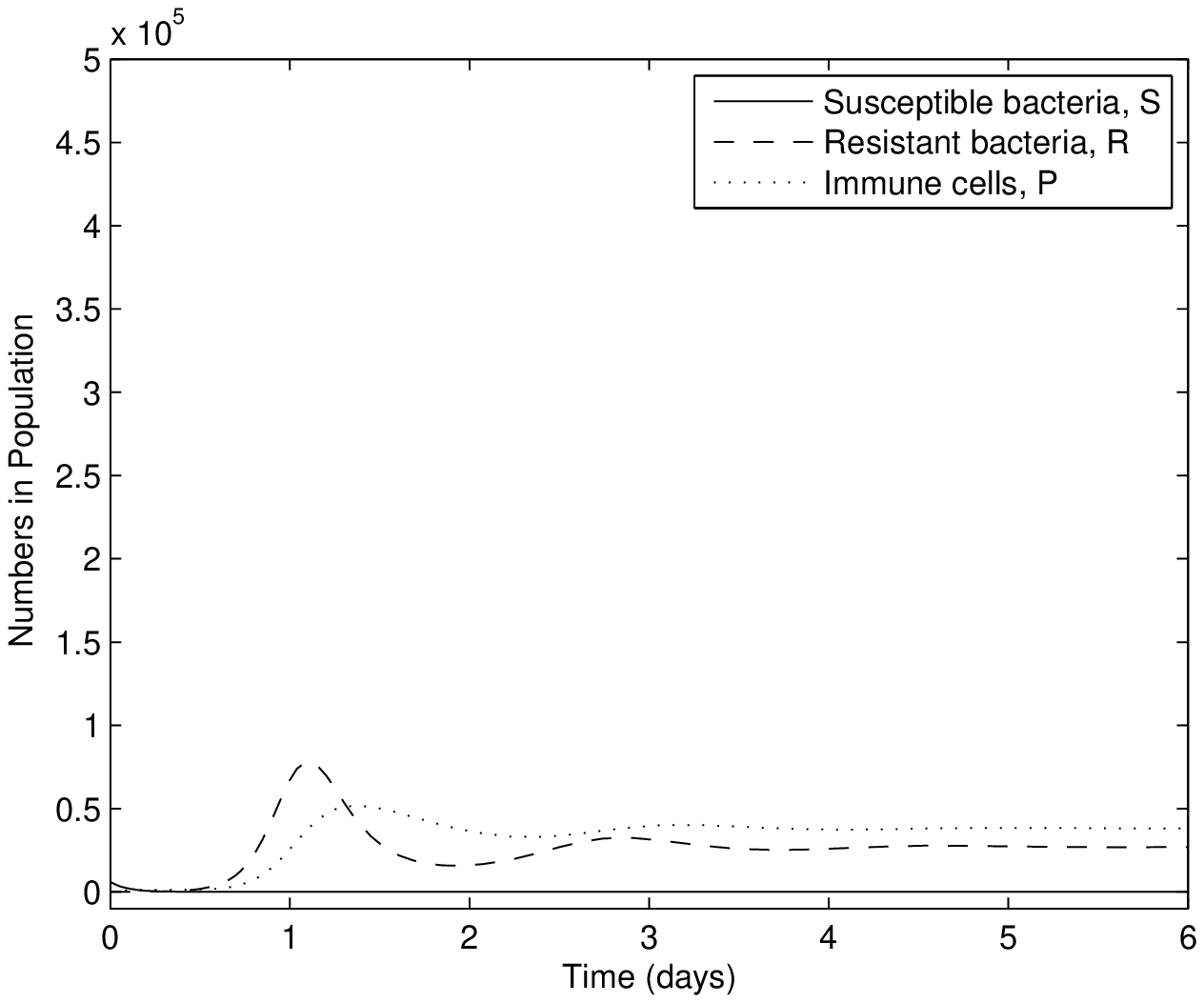}
\caption{}
\label{InitialSimulationModel1}
\end{subfigure}
\caption{Numerical solutions to \eqref{Model1Antibiotic}-\eqref{Req} with (a) no treatment ($A(0)=A^*(0)=0$), (b) one dose of antibiotic ($A(0)=4, A^*(0)=0$), (c) repeated dosing of antibiotic ($A(\tau)=4$ for $\tau=0.5p$, $p\in{0,1,2,...}$ and $A^*(0)=0$) and (d) constant antibiotic level ($A(0)=4, A^*(0)=0, \alpha=0$).
With no treatment, our parameter choice reflects the situation where the immune system cannot clear the infection and a dominant population of antibiotic-resistant bacteria emerge.
With treatment, susceptible bacteria can be cleared but some antibiotic-resistant bacteria persist.
Note that here and in subsequent figures, we have adjusted the $y$-axes to enable visualisation of any variables tending to zero.}
\label{InitialSimulationsResistance}
\end{figure}

We consider a scenario where the bacterial population at the site of infection is composed predominantly of an antibiotic-susceptible subpopulation and a minor, resistant subpopulation which has arisen through cross-contamination of the site with a resistant population introduced from an environmental source (as is often the case during hospital-outbreaks) \cite{Das:2002,Fanci:2009}.

Firstly, we establish that our parameter set predicts the emergence of antibiotic-resistant bacteria in an infection when initially they exist in far smaller numbers than susceptible bacteria (Figure \ref{InitialSimulationNoAntibiotic}) but ultimately dominate the infection as a result of conjugation.
The parameter choice also reflects an infection that the immune system alone cannot clear (hence the need for treatment).
Due to acquisition of resistance via horizontal evolution from plasmid transfer the resistant bacteria become dominant even without the selective pressure of an antibiotic.

Antibiotics can either be administered in doses ($\alpha\neq0$) or continuously ($\alpha=0$).
Our  model predicts that under antibiotic dosing all susceptible bacteria may be eradicated from the infection, however, as the antibiotic degrades out of the system the population of resistant bacteria increases exponentially (Figure \ref{DegradingAntibioticModel1} and \ref{Model1DosingSimulation}). 
Upon a new dose of antibiotics being administered, the level of resistant bacteria drops slightly (due to the allowance of only partial resistance), subsequently quickly rising when the antibiotic degrades out of the system.
Next, we consider constant antibiotic administration (e.g. via intravenous therapy, Figure \ref{InitialSimulationModel1}).
Susceptible bacteria are eliminated and the resistant bacteria reach a lower (but still high) steady state than in the absence of any drug.
Thus the model reflects the type of infection in which we are most interested: one which cannot be cleared by either the immune system or antibiotics and in which antibiotic-resistance is a primary concern.
For the remainder of this study we consider only constant antibiotic concentration for mathematical simplicity.

\paragraph{Antibiotic treatment: parameter analysis \\}

\begin{figure}
\centering
\begin{subfigure}[b]{0.49\textwidth}
							 \includegraphics[width=\textwidth]{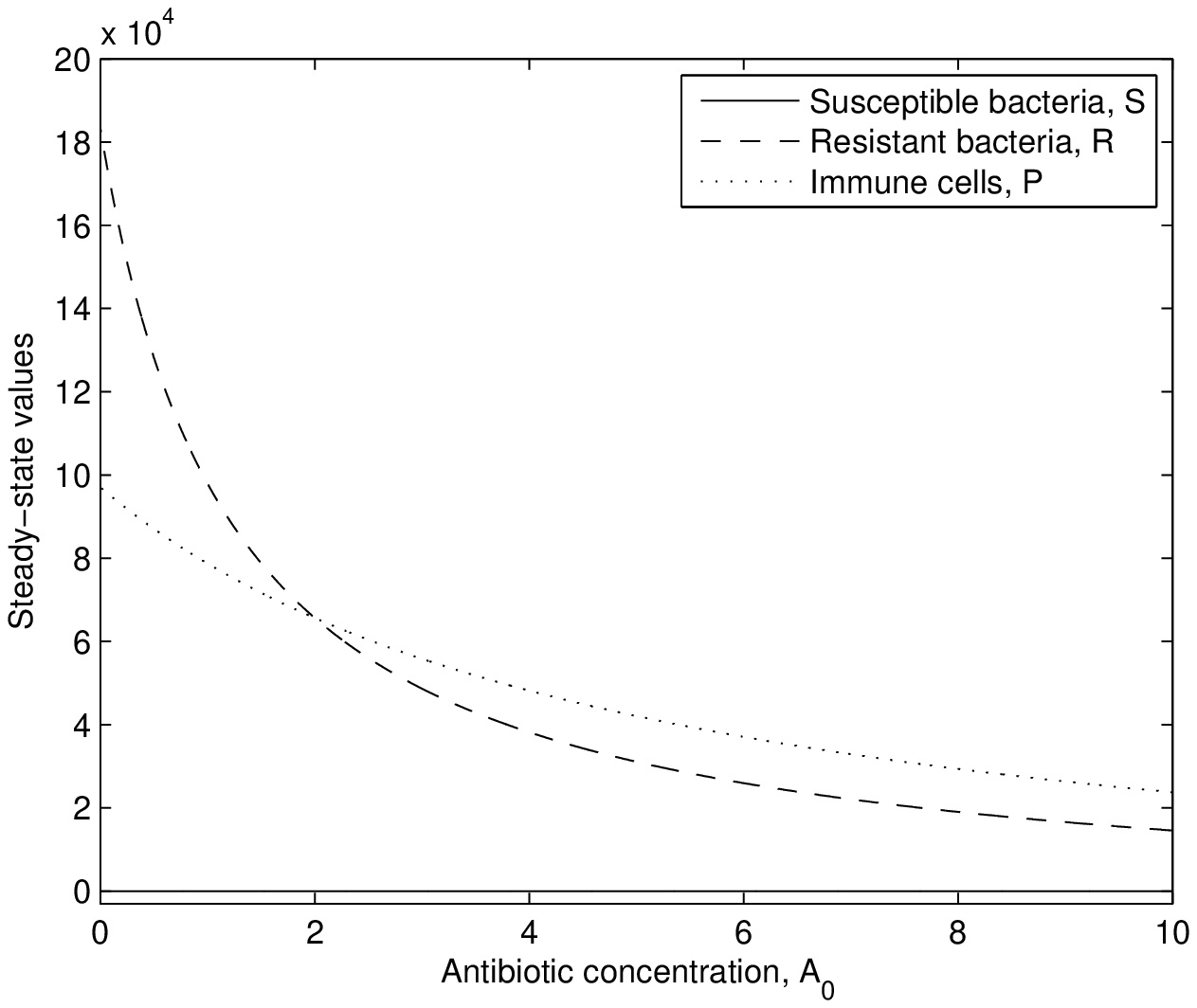}
          \caption{}
          \label{A0_ss_new}
   \end{subfigure}       
\begin{subfigure}[b]{0.49\textwidth}
\includegraphics[width=\textwidth]{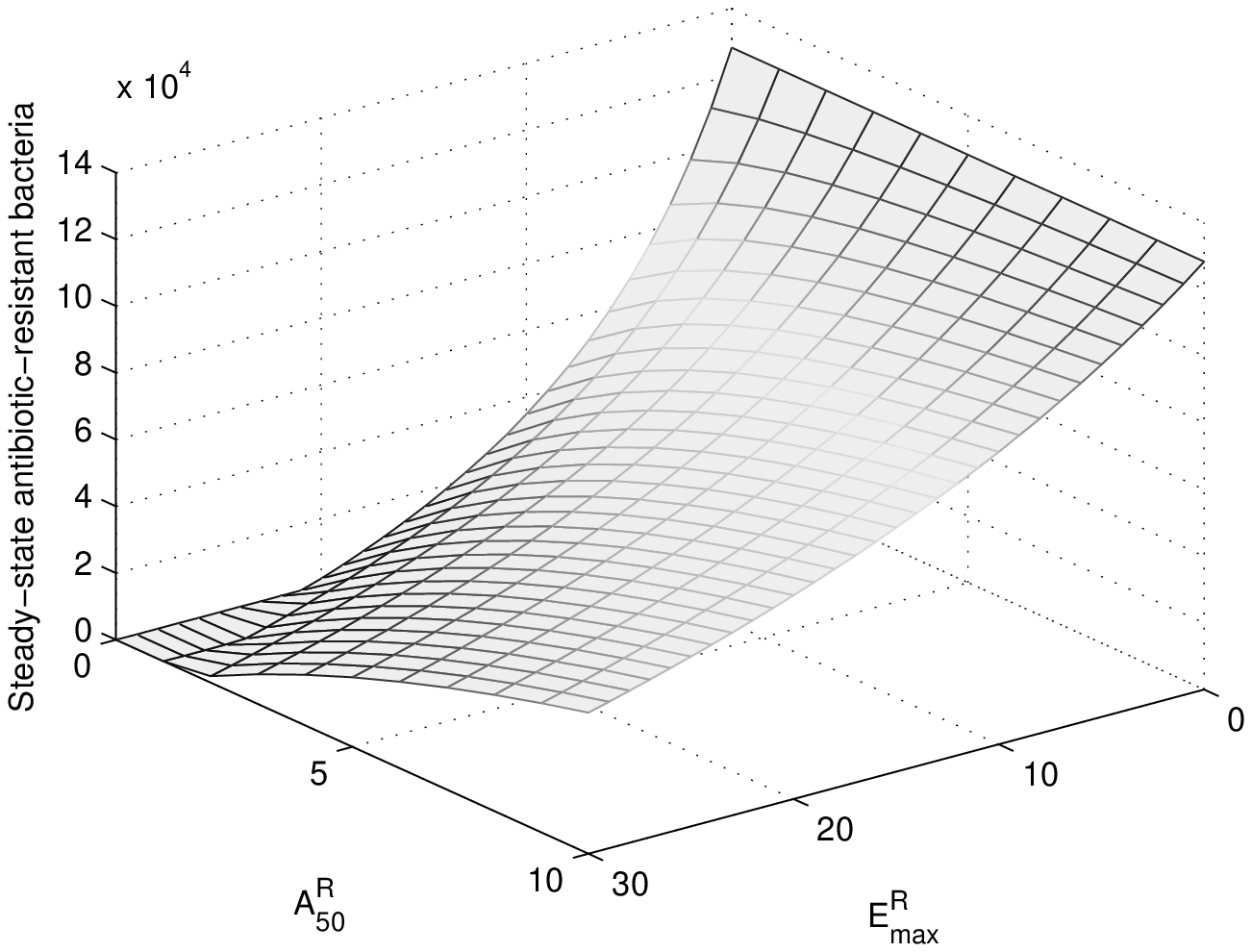}
\caption{}
\label{ERvsARSurfacePlot}
\end{subfigure}          
\caption{ Steady-state values of antibiotic-resistant bacteria for varying (a) $A(0)$, (b) $A_{50}^R$ and $E_{\text{max}}^R$ (the last two simultaneously).
In (a) we plot also the numbers of antibiotic-susceptible bacteria to illustrate that they are cleared from the infection.
Antibiotic-resistant bacteria will only be cleared by antibiotics (in combination with the immune system) if resistance is especially low ($A_{50}^R\to0$ and $E_\text{max}^R\to\infty$).}         
\end{figure}

Evidently, the amount of antibiotic administered will have a large effect on the final relative population levels of antibiotic-susceptible and -resistant bacteria (Figure \ref{A0_ss_new}).
Though some resistant bacteria persist regardless of the dosage, as a result of the default parameter set representing only partial resistance, the bacterial load can be lowered by increasing the quantity of antibiotic administered (though $A_0$ must remain in a clinically realistic range).

Partial resistance is governed by the two parameters $E^R_{\text{max}}$ and $A^R_{50}$ (the antibiotic's maximum killing rate for resistant bacteria and the antibiotic concentration for half the maximum effect on resistant bacteria, respectively). 
The limits $E^R_{\text{max}} \to 0$ or $A^R_{50} \to \infty$ yield complete resistance, and in Figure  \ref{ERvsARSurfacePlot} we see what happens when these parameters are varied in combination.
The sensitivity of the bacteria to a specific antibiotic determines the bacterial load that persists in an infection treated by antibiotics: reducing the resistance level lowers this number.

Though the results in this section so far are in no way surprising, they illustrate the power of adopting a modelling approach to predict the outcome of a particular infection: if the infecting strain can be isolated and the relevant parameters can be determined, it is possible to predict the number of bacteria remaining at an infection site and therefore identify an optimal treatment strategy.

\begin{figure}
\centering
\begin{subfigure}[b]{0.485\textwidth}
\includegraphics[width=\textwidth]{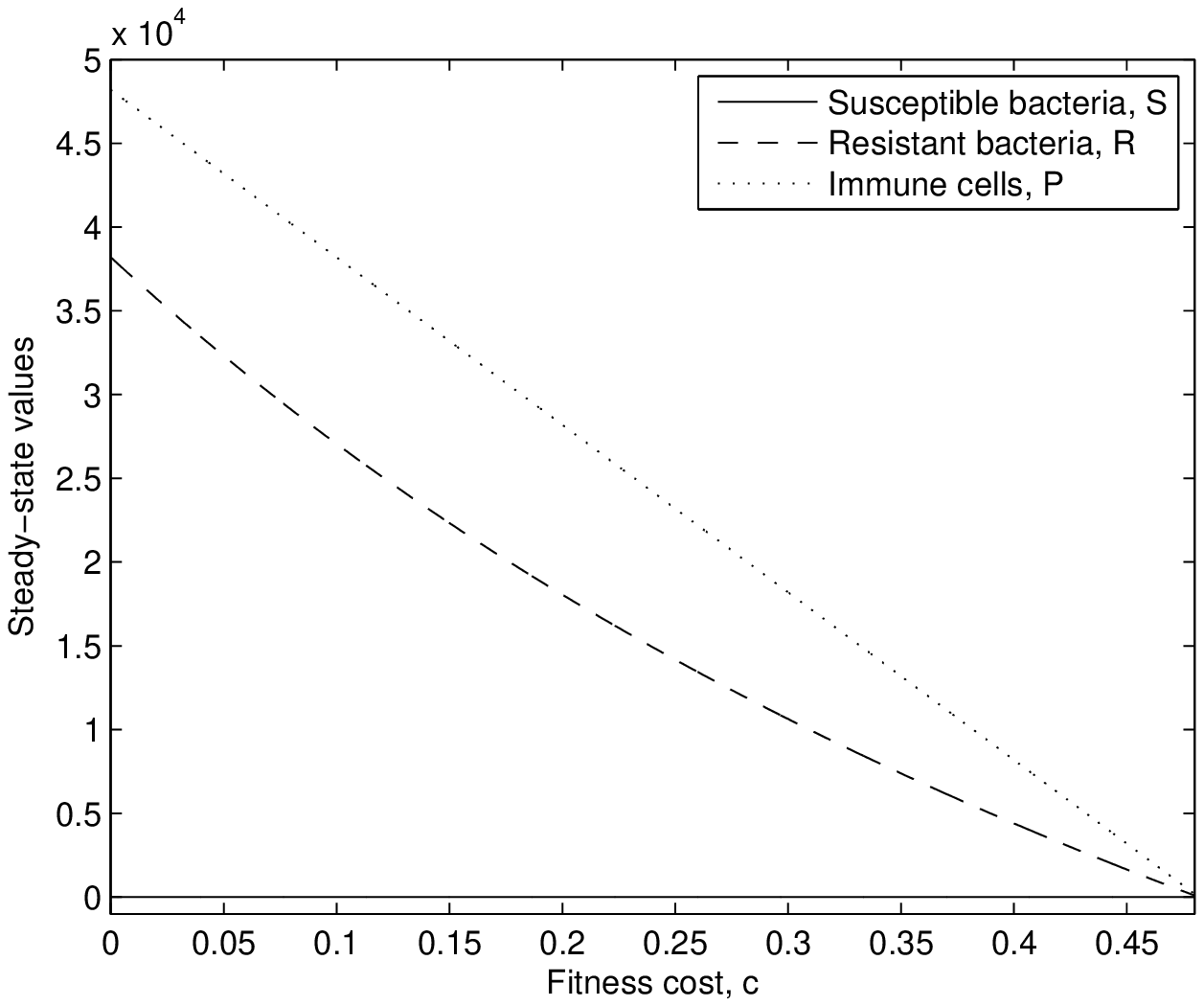}
\caption{}
\label{VaryingcModel1}
\end{subfigure}
\quad
\begin{subfigure}[b]{0.485\textwidth}
\includegraphics[width=\textwidth]{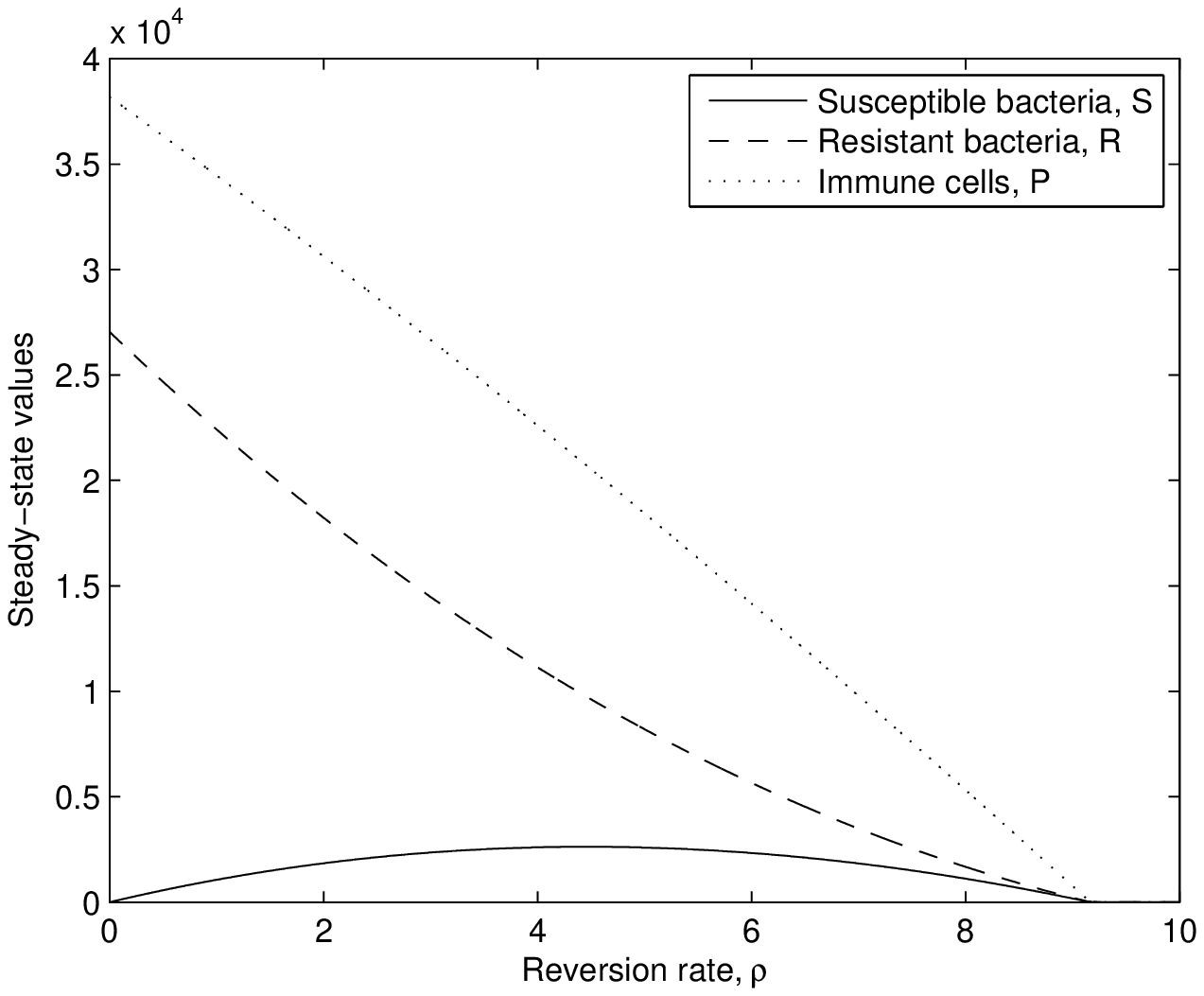}
\caption{}
\label{VaryingRhoModel1}
\end{subfigure}
\caption{Steady-state values of $S$, $R$ and $P$ in response to varying (a) fitness cost, $c$, and (b) reversion rate, $\rho$.
Sufficiently high $c$ results in both subpopulations dying out (the susceptible bacteria as a result of the antibiotic and the resistant bacteria because the fitness cost renders them no longer viable) while there exists a range of $\rho$ where \textit{both} subpopulations can persist in the long-term in spite of treatment by antibiotic.}
\label{VaryingCandRhoModel1}
\end{figure}

We recall that a fitness cost can be incurred as a consequence of the bacteria bearing the plasmid that confers antibiotic resistance.
It has been shown that, depending on the plasmid in question, the fitness cost incurred can vary from 0.09 to 0.25 \cite{Subbiah:2011aa} (as a proportion of the non-plasmid-bearing growth rate), and that, over time, this cost can reduce, sometimes eventually disappearing altogether \cite{Dionisio:2005aa}. 
Thus, it is important to look at the effect of varying the cost of resistance on the whole infection (Figure \ref{VaryingcModel1}).

We observe that, as expected, a decrease in fitness cost increases the levels of resistant bacteria, and thus the level of immune cells (which are recruited in response to the presence of bacteria), at the infection site. 
As $c \to 0$, the plasmid is no longer exerting any significant metabolic cost on the bacteria and thus they are in a stronger position when faced with antibiotic pressure. Similarly, as $c$ increases, we see the bacteria unable to compete against this antibiotic pressure and once $c > 0.48$ (for our default parameter set), the benefit of resistance no longer outweighs the costs incurred and the antibiotic-resistant subpopulation dies out in the long term.

We also find that variations to the reversion rate, $\rho$, the rate at which a plasmid-bearing bacteria loses its plasmid \cite{ Webb:2005aa, Imran:2006aa} can cause significant changes in the dynamics of the system (Figure \ref{VaryingRhoModel1}). 
There exists a range of $\rho$ where populations of both antibiotic-resistant and -susceptible bacteria are maintained in the population at steady state despite the presence of antibiotic: though the antibiotic kills the susceptible bacteria the resistant bacteria rejuvenates the susceptible population via plasmid loss. 
However, a sufficiently fast rate of plasmid loss results in both populations dying out: the resistant population all revert back to plasmid free susceptible bacteria leaving no plasmid-bearing bacteria to rejuvenate the population once the antibiotic has taken effect. 
In reality we expect this reversion rate to be very low so the situation whereby only resistant bacteria remain in the population is most likely \cite{Sorensen:2005aa}.\\

\paragraph{Anti-virulence drugs ($\mathbf{A(0)=0}$ $\mathbf{\mu}$g/ml, $\mathbf{A^*(0)=4}$ $\mathbf{\mu}$g/ml)\\}

We consider how effective anti-virulence drugs alone can be in treating a bacterial infection without antibiotics.
\begin{figure}
\centering
\begin{subfigure}[b]{0.485\textwidth}
\includegraphics[width=\textwidth]{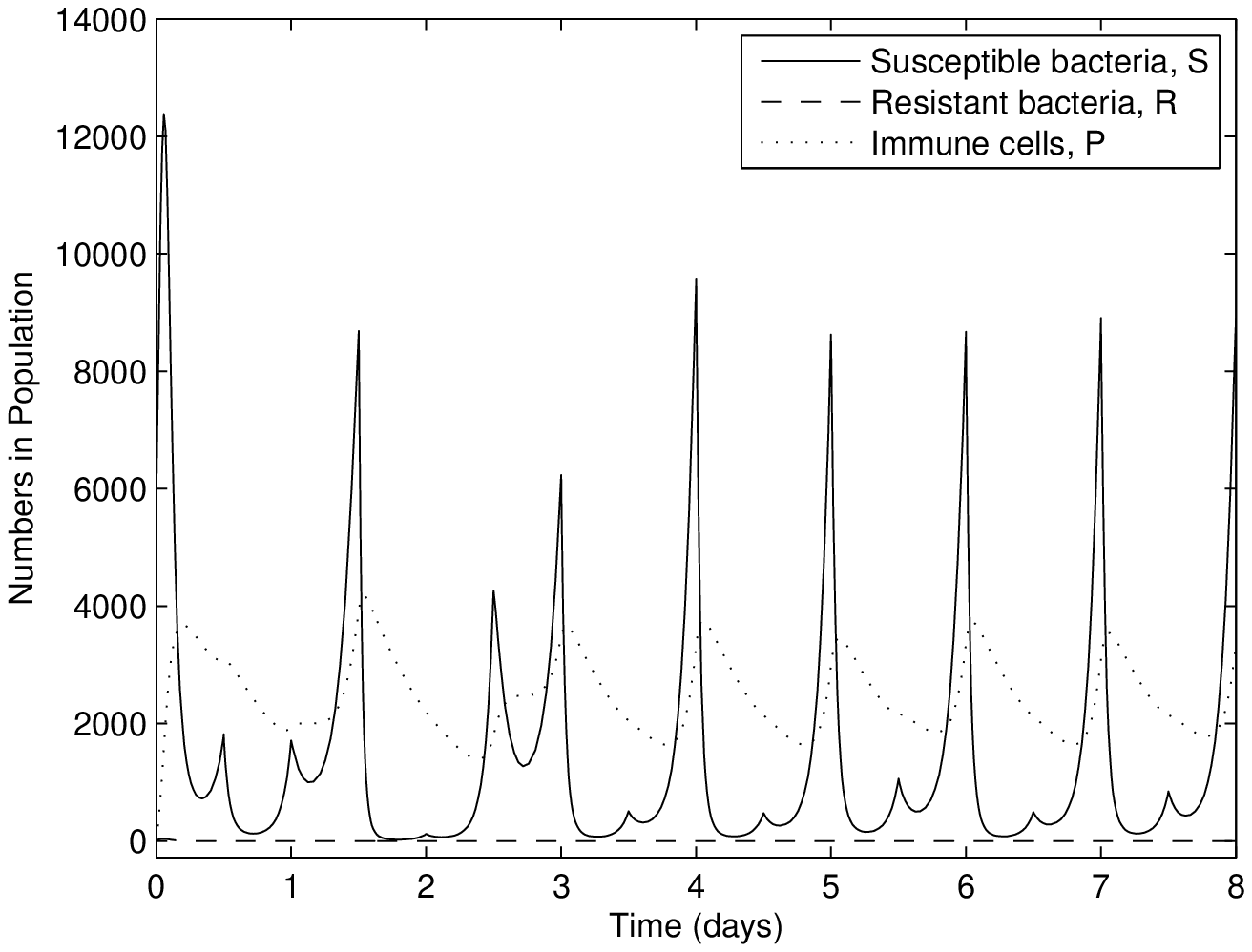}
\caption{}
\label{AV4dosing}
\end{subfigure}
\quad
\begin{subfigure}[b]{0.485\textwidth}
\includegraphics[width=\textwidth]{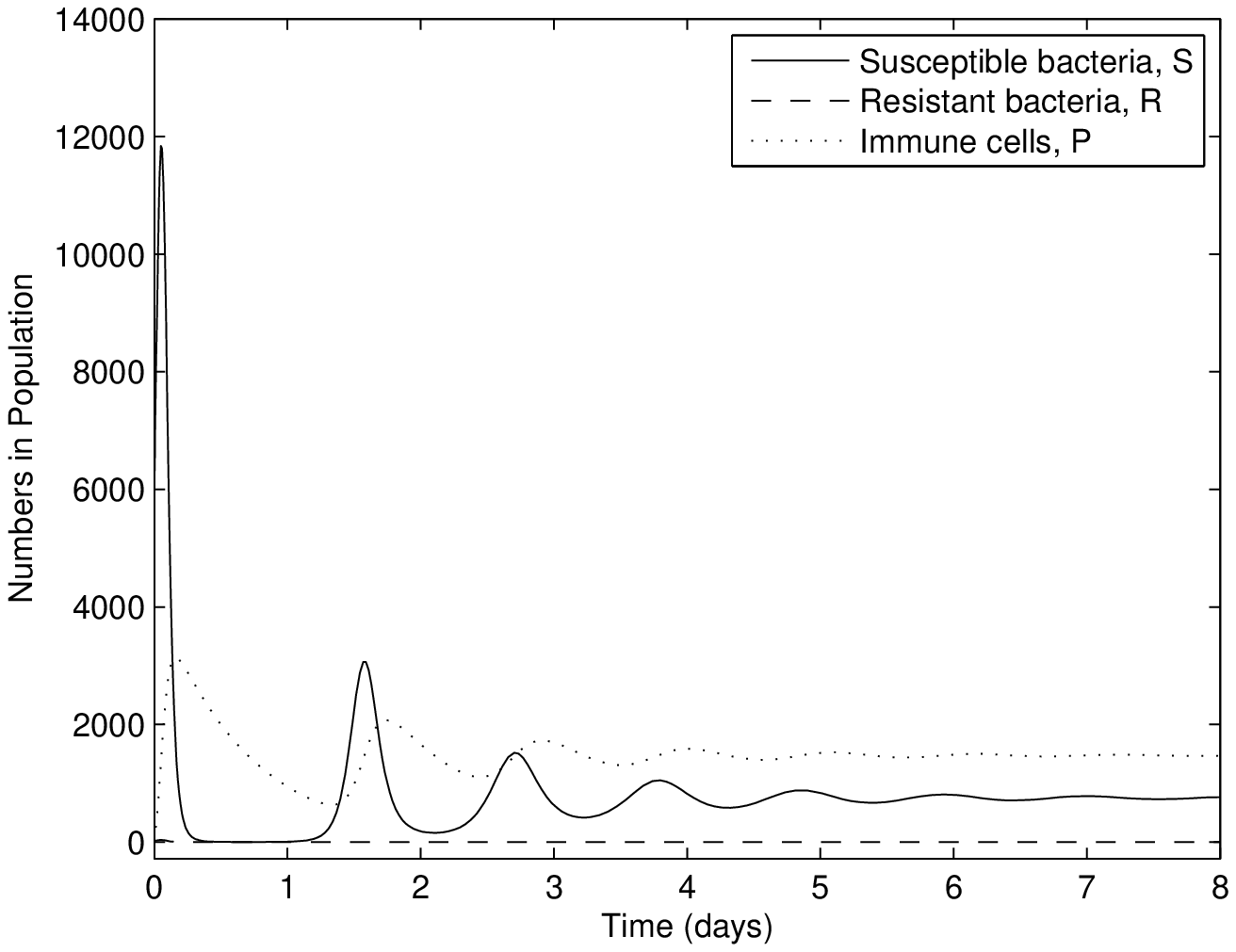}
\caption{}
\label{Astar4c0p1}
\end{subfigure}
\caption{Numerical solution to \eqref{Model1Antibiotic}-\eqref{Req} when the infection is treated only by anti-virulence drugs ($A(0)=0,A^*(0)=4$) (a) with doses every 0.5 days ($\kappa=3.6$) and (b) continuous treatment ($\kappa=0$).
Though the antibiotic-resistant bacteria are quickly cleared from the infection, a subpopulation of antibiotic-susceptible bacteria persist because they grow at a faster rate than the antibiotic-resistant bacteria.
The reduced number of immune cells present at the infection site as a result of the treatment lowering the total bacterial load leaves the immune response incapable of clearing these remaining bacteria.
For the remainder of this study we consider only constant anti-virulence drug for simplicity.
}\label{InitialSimulationModel2}
\end{figure}
Under our default parameter set (Figure \ref{InitialSimulationModel2}) the anti-virulence drug eliminates the antibiotic-resistant bacteria (this subpopulation is weaker due to the fitness cost imposed on them via the presence of the plasmid).
However, as opposed to an infection treated only by antibiotics where antibiotic-resistant bacteria persisted, the anti-virulence drug fails to eliminate all the antibiotic-susceptible bacteria.
Encouragingly though, the number of bacteria remaining in the system is much lower than the number remaining after antibiotic treatment.

Despite the anti-virulence drug making the bacteria more vulnerable to the immune system there is still a limit to how effective the immune response can be and hence why some bacteria remain at the infection site. 
In modelling recruitment of the immune cells to the infection site, we imposed a recruitment rate, $\beta$, proportional to the number of bacteria present at the infection site, and a carrying capacity, $P_{\text{max}}$. 
As such, the presence of immune cells is limited by the presence of bacteria. Subsequently, as the drug starts to take effect the bacterial population diminishes, shortly followed by a decline in phagocyte population. Since this treatment is based on enabling the immune system to clear the infection, as the immune cell levels decrease so does the efficacy of the drug; hence we see a possible reason for the persisting population of susceptible bacteria.

Thus although anti-virulence drugs are capable of dealing with the presence of antibiotic-resistant bacteria if they grow more slowly than antibiotic-susceptible bacteria, it may not be an effective treatment for the infection as a whole. 
In the following parameter analysis we consider whether anti-virulence drugs would ever be capable of clearing an infection or whether they might be more suitable for infection prevention.

\paragraph{Anti-virulence drugs: parameter analysis\\}

\begin{figure}
\centering
                \includegraphics[width=0.65\textwidth]{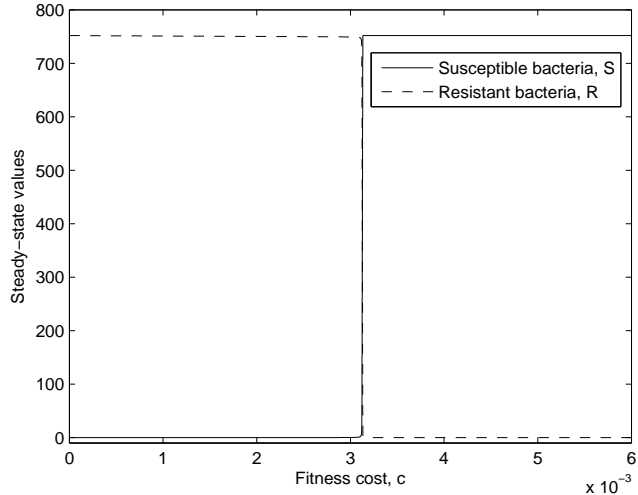}
          \caption{Steady state values of $S$ and $R$ for varying values of the cost of resistance, $c$, with $A(0)=0, A^*(0)=4$ and all other parameters as given in Table \ref{Table1}. We see how, at an approximate fitness cost of $c=0.0031$, the system exhibits a switch in behaviour and the steady state values for antibiotic-resistant and antibiotic-susceptible bacteria are reversed, owing to the susceptible bacteria being the weaker strain when the fitness cost for bearing the plasmid is sufficiently small. }\label{VaryingCModel2}
\end{figure}

Since the fitness cost is something that varies between bacteria and resistance type, we again consider variations to the parameter $c$.
This parameter effectively controls a competition for dominance between the antibiotic-resistant and -susceptible bacteria. 
Indeed, looking at Figure \ref{VaryingCModel2}, we see striking behaviour in the regime $0<c<0.006$. 
For $0<c < 0.0031$ the anti-virulence drug eliminates the population of antibiotic-susceptible bacteria with a level of antibiotic-resistant bacteria persisting in the system. 
With such a small resistance cost the susceptible bacteria are the more vulnerable strain and thus, when in competition with the antibiotic-resistant bacteria, the susceptible bacteria lose out and experience the full effect of the anti-virulence drug. 
However, when $c > 0.0031$, the cost of antibiotic-resistance does not outweigh the benefits from plasmid transfer and the plasmid-bearing bacteria are the more vulnerable strain and are hence eliminated.
This highlights the importance of understanding the specific bacteria in question when choosing a treatment strategy for an infection.

A crucial aspect of an anti-virulence drug is its dependence on the immune system to be successful: it may not be able to fully clear an infection under our default parameter set (Figure \ref{InitialSimulationModel2}).
Both the natural removal rate, $\psi$, and the baseline immune response, $\gamma$, are not only very difficult to estimate, they are also patient- and site- specific. 
Along with inherent individual differences in patients, we also see variation within a single host depending on their current health. Since the main cause of concern for antibiotic resistance is in nosocomial infections \cite{Alanis:2005aa} patients may well be ill, immunosupressed or debilitated before the onset of infection. 

\begin{figure}
\centering
\begin{subfigure}[b]{0.485\textwidth}
\includegraphics[width=\textwidth]{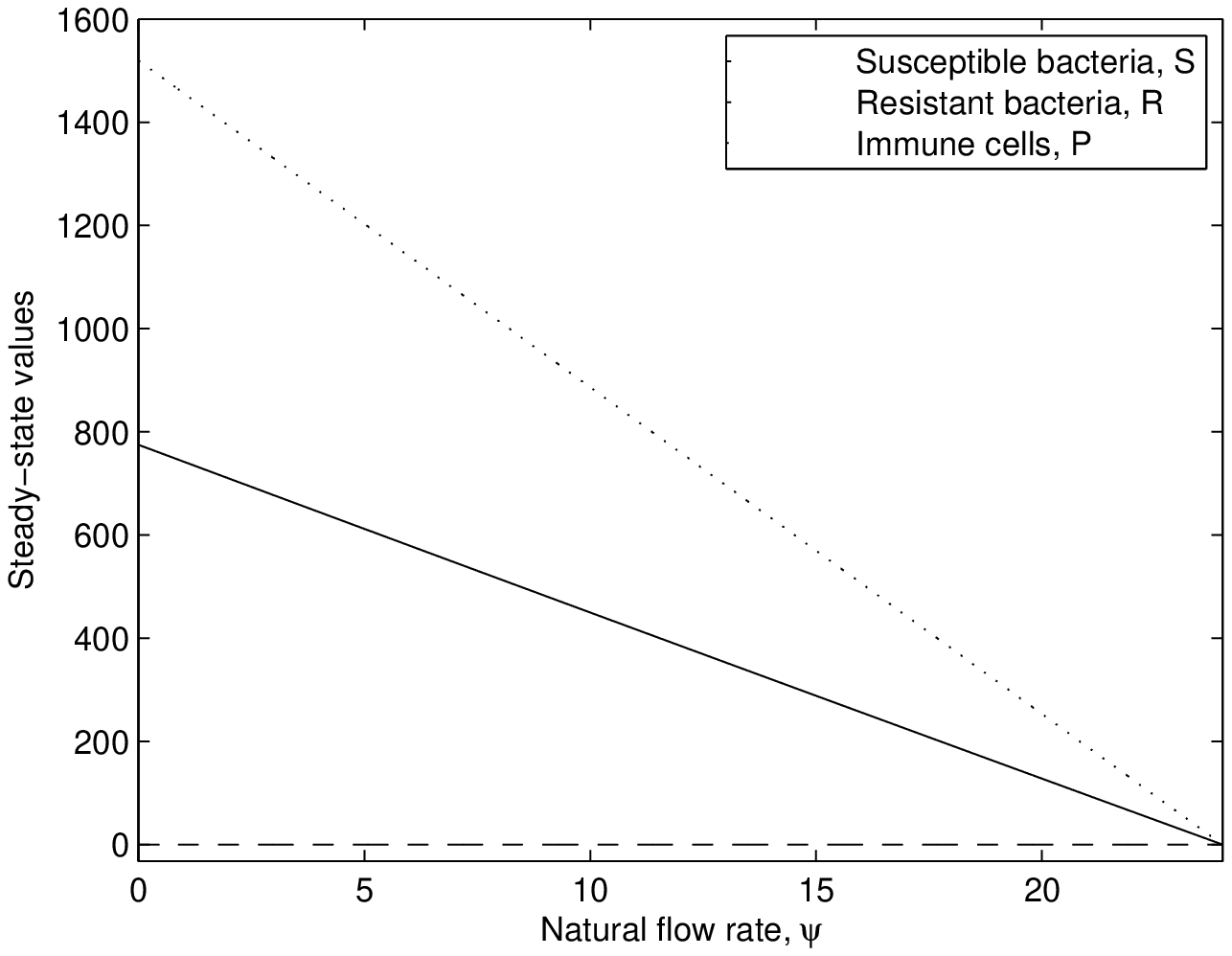}
\caption{}
\label{VaryingPsiModel2}
\end{subfigure}
\quad
\begin{subfigure}[b]{0.485\textwidth}
\includegraphics[width=\textwidth]{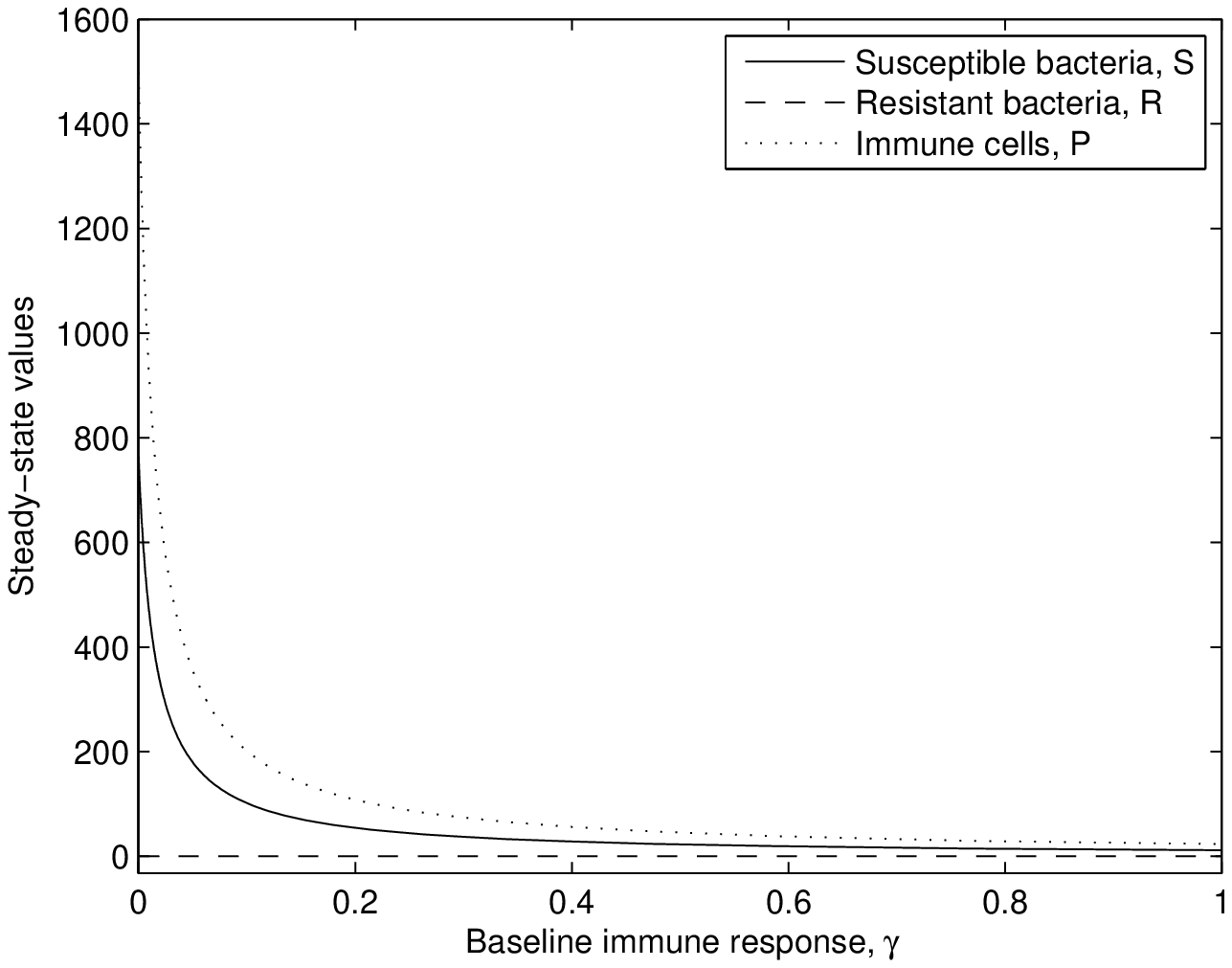}
\caption{}
\label{VaryingGammaModel2}
\end{subfigure}
\caption{Steady-state value of susceptible bacteria and phagocyte levels when (a) natural clearance ($\psi$) and (b) baseline immune response ($\gamma$) are varied.
The bacteria can be cleared from the infection but only when these host-specific parameters are sufficiently strong.}
\label{HostParametersModel2}
\end{figure}

Figure \ref{HostParametersModel2} shows the effect of varying these host-specific parameters on the anti-virulence treatment. 
There exist circumstances when anti-virulence drugs can be completely effective, however these occur when both the baseline immune response and natural flow past the infection site are very high. 
Thus it is possible that anti-virulence drugs may only be successful in either treating or preventing infection in a healthy individual.
To test infection prevention in a model, a more sophisticated set of equations that capture a specific mechanism of action by an anti-virulence drug, as opposed to our deliberately general conceptual model, is required.
This will be developed in future work.

In all cases examined here the strain of resistant bacteria are eliminated, due to their reduced fitness making them more vulnerable to the treatment. 
Thus, for an existing infection requiring treatment, given that only susceptible bacteria remain, we next examine the option of combining treatment by anti-virulence drugs with conventional antibiotics.

\paragraph{Combination therapy: antibiotics and anti-virulence drugs ($\mathbf{A(0)=4}$ $\mathbf{\mu}$g/ml, $\mathbf{A^*(0)=4}$ $\mathbf{\mu}$g/ml) \\}

\begin{figure}
\centering
\begin{subfigure}[b]{0.485\textwidth}
\includegraphics[width=\textwidth]{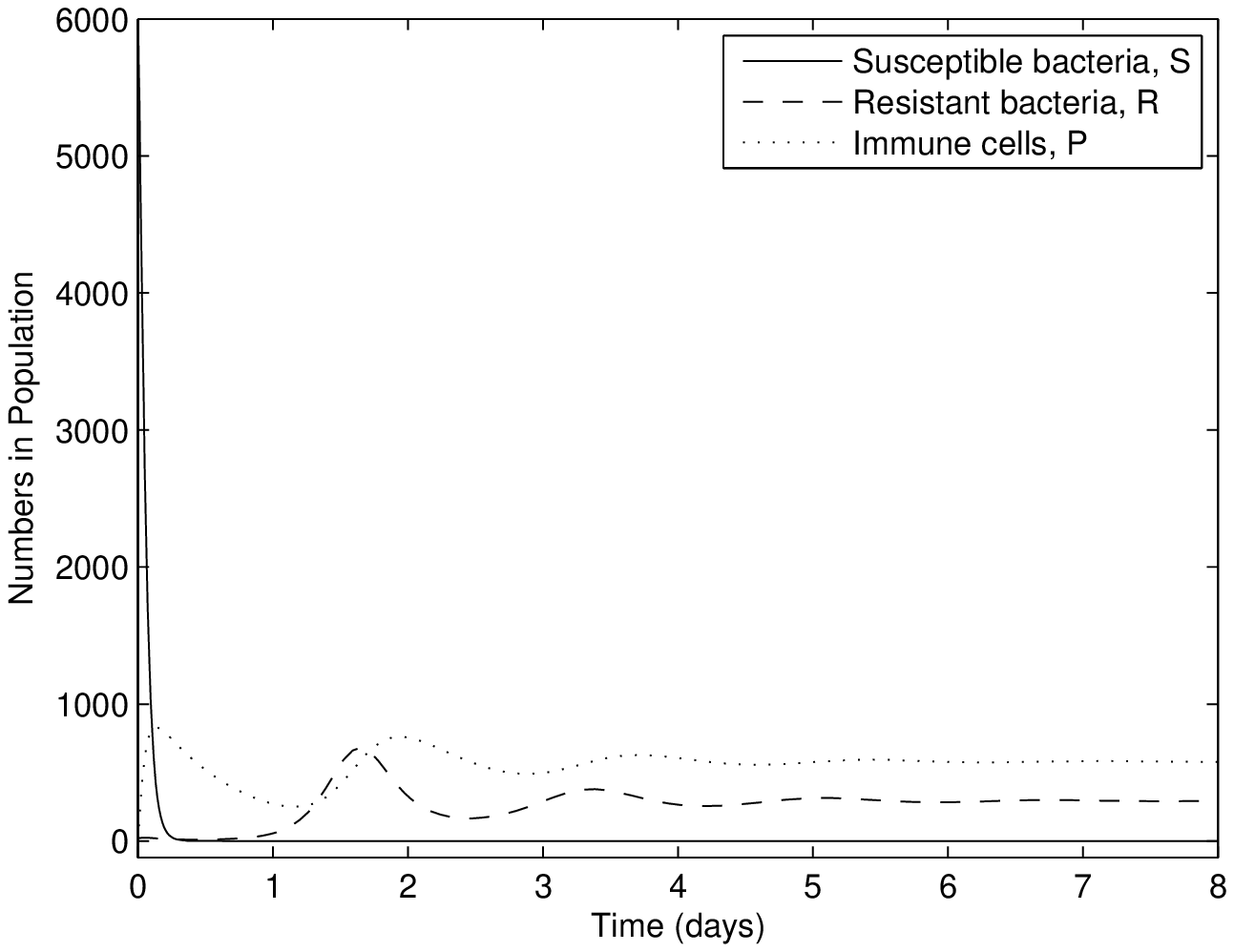}
\caption{}
\label{A4Astar4c0p1}
\end{subfigure}
\quad
\begin{subfigure}[b]{0.485\textwidth}
\includegraphics[width=\textwidth]{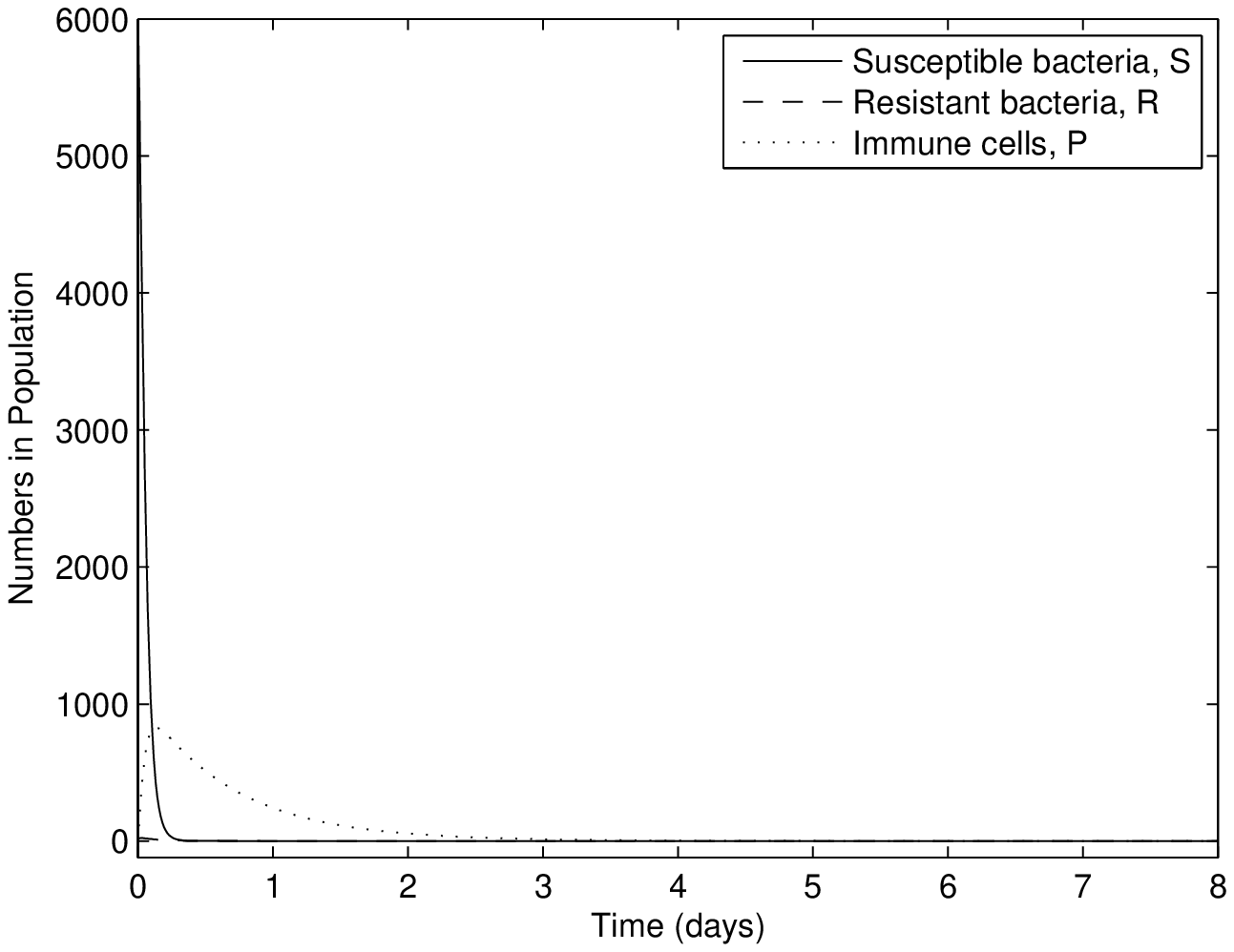}
\caption{}
\label{A4_Astar4_c0p27}
\end{subfigure}

          \caption{Simulation of combination therapy ($A(0)=A^*(0)=4, \alpha=\kappa=0$) with (a) $c=0.1$ and (b) $c=0.27$.The combined treatments are only successful if the fitness cost associated with antibiotic resistance is sufficiently high.   }\label{CoupledSameTime}
\end{figure}

Our results so far suggest that anti-virulence drugs would rid an infection site of antibiotic-resistant but not antibiotic-susceptible bacteria, so it would be expected that the addition of antibiotics to this treatment would fully clear the infection.
However, somewhat surprisingly, Figure \ref{CoupledSameTime}a indicates that this is not necessarily the case  (even under the same parameter set): though susceptible bacteria are cleared, a number of resistant bacteria persist if the fitness cost associated with antibiotic resistance is sufficiently low.
This may be due to the dynamics of the immune cells: as the introduction of both drugs rapidly decreases the population of susceptible bacteria we also see a decline in the number of immune cells. 
Due to the dependence of the anti-virulence drug on the immune system, this decline may subsequently stunt the effectiveness of the drug, allowing the antibiotic-resistant population to build up to a level with which the immune system then cannot compete. 
In Figure \ref{CoupledSameTime}b, however, we see that raising the fitness cost to $c=0.27$  can result in both subpopulations being eliminated due to the additional weakness imposed on the antibiotic-resistant bacteria (we note that this is just outside the possible range for $c$ suggested in \cite{Subbiah:2011aa}).

\begin{figure}
\centering
\begin{subfigure}[b]{0.485\textwidth}
\includegraphics[width=\textwidth]{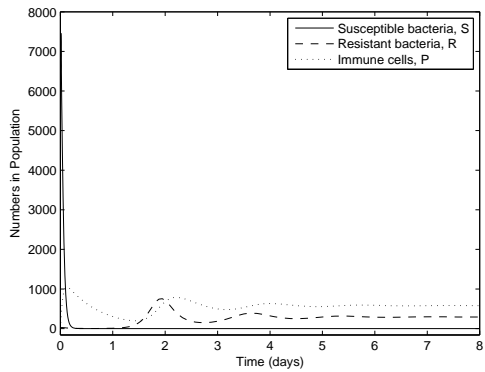}
\caption{}
\label{AVthenAB4a}
\end{subfigure}
\quad
\begin{subfigure}[b]{0.485\textwidth}
\includegraphics[width=\textwidth]{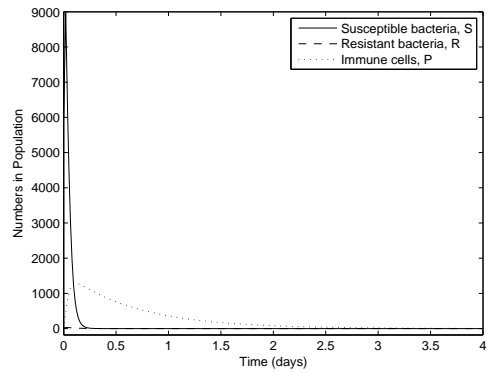}
\caption{}
\label{AVthenAB5a}
\end{subfigure}
\caption{Results of administering an anti-virulence drug to the infection site, followed by an antibiotic after (a) 0.01 days and (b) 0.02 days when $c=0.1$.
If the delay between drugs is sufficient, the anti-virulence drug is able to eradicate the antibiotic-resistant bacteria, allowing the antibiotic to deal with the remaining susceptible bacteria.}
\label{CoupledTimeDelay}
\end{figure}

Excitingly, imposing a time delay on the administration of one of the drugs can allow combination therapy to be successful regardless of the fitness cost: see Figure \ref{CoupledTimeDelay}. 
\begin{figure}
\centering
\begin{subfigure}[b]{0.485\textwidth}
\includegraphics[width=\textwidth]{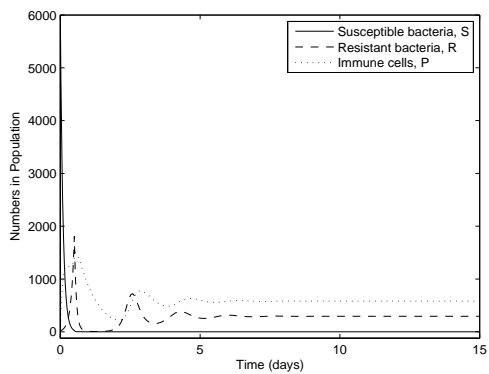}
\caption{}
\label{ABthenAV8a}
\end{subfigure}
\quad
\begin{subfigure}[b]{0.485\textwidth}
\includegraphics[width=\textwidth]{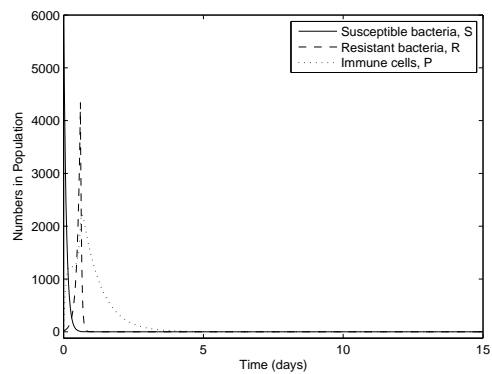}
\caption{}
\label{ABthenAV9a}
\end{subfigure}
\caption{Results of administering an antibiotic to the infection site, followed by an anti-virulence drug after (a) 0.5 days and (b) 0.6 days with $c=0.1$.
The required delay between drugs for clearance of the infection is longer if the drugs are administered in this order than vice versa (compare with Figure \ref{CoupledTimeDelay}).
}
\label{CoupledTimeDelayABthenAV}
\end{figure}
When $c=0.1$, the infection can be fully cleared if the antibiotic is administered 0.02 days (i.e. roughly 30 minutes) after the anti-virulence drugs, though the bacterial load in this time delay will greatly increase.
Reversing the treatment order is also successful, see Figure \ref{CoupledTimeDelayABthenAV}.
Here the delay is longer (and hence clearing the infection takes longer) but the increase in bacterial load before addition of the second drug is lower in this instance.
Thus, we see that in combining two treatments, one which is more effective towards resistant bacteria and one which favours susceptible bacteria, we can effectively eliminate all bacteria, provided a time delay is incorporated to enable both drugs to have full effect.
%We thus highlight the usefulness of mathematical modelling in designing treatment strategies.

\begin{figure}
\centering
\begin{subfigure}[b]{0.485\textwidth}
\includegraphics[width=\textwidth]{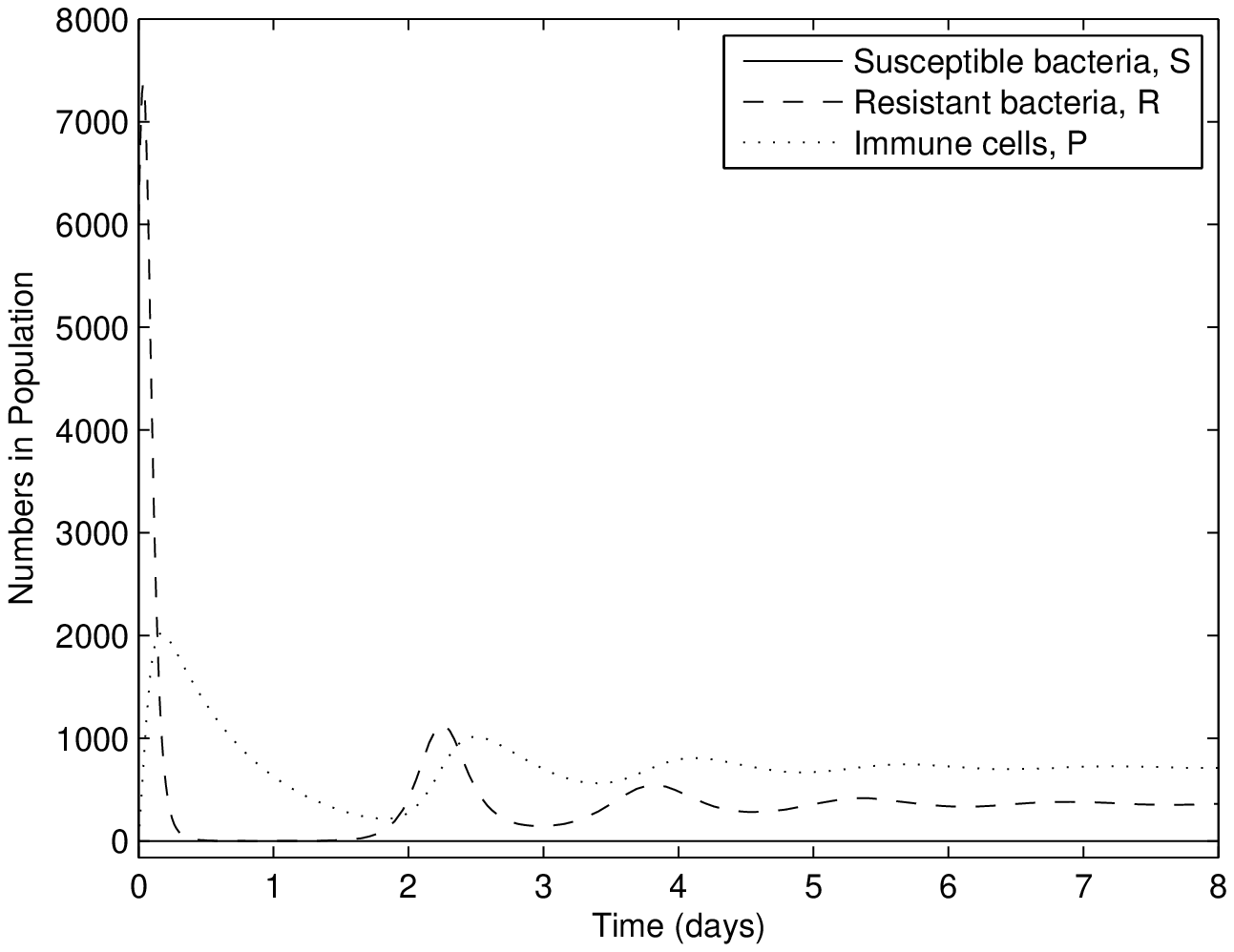}
\caption{}
\label{FullRes2}
\end{subfigure}
\quad
\begin{subfigure}[b]{0.485\textwidth}
\includegraphics[width=\textwidth]{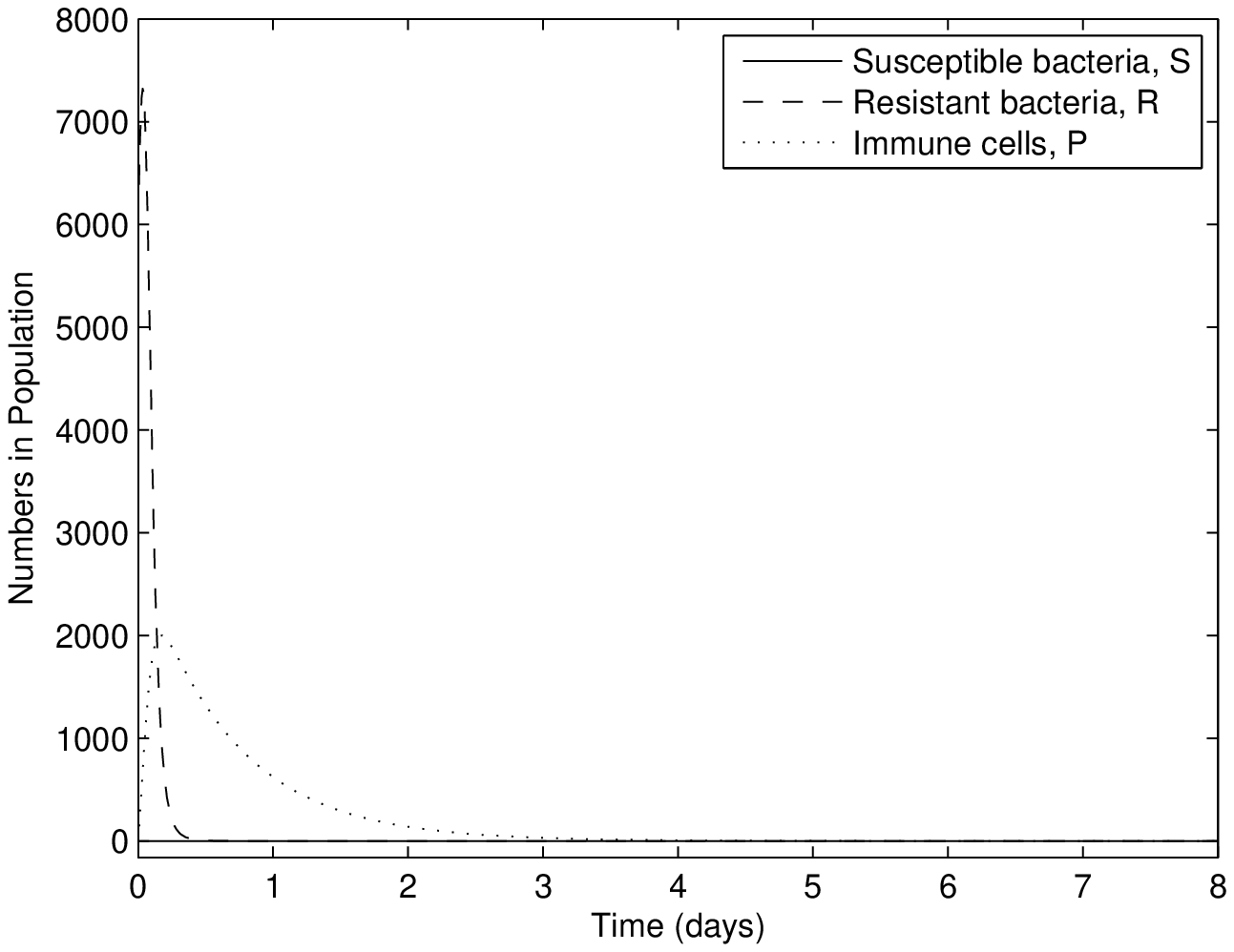}
\caption{}
\label{FullRes3}
\end{subfigure}
\caption{An infection consisting solely of (partially) antibiotic-resistant bacteria ($S(0)=0, R(0)=6000$) treated by both antibiotics and anti-virulence drugs simultaneously with (a) $c=0.01$ and (b) $c=0.02$. The combination therapy is capable of clearing the infection, but only if the fitness cost of antibiotic resistance is sufficiently high.}
\label{FullRes}
\end{figure}

So far, all our simulations have focused on an infection where the antibiotic-resistant population are initially in the minority. In Figure \ref{FullRes} we consider the effect of this combination therapy on an infection consisting entirely of antibiotic-resistant bacteria ($S(0)=0, R(0)=6000$).
Interestingly, though neither drug in isolation will clear the infection, the administration of both drugs with no time delay between them can clear the infection provided $c>0.01$ for our default parameter set.
Thus combining antibiotics and anti-virulence drugs can even be successful in treating infections consisting solely of antibiotic-resistant bacteria (remembering that we are considering the bacteria to be partially resistant to the antibiotics).
In addition, if we consider the scenario where antibiotic-resistance has occurred via spontaneous chromosomal mutation (hence $\lambda=\rho=0$) and no fitness cost is therefore incurred ($c=0$), delayed combination therapy again works, this time with delays of 0.03 days if the anti-virulence drug is administered first and 0.5 days if vice versa (data not shown), again underlining the need to fully understand the infection-type to design the treatment strategy correctly.

\section{Discussion}

Bacterial resistance to antibiotics is an increasing problem in today's society, more specifically in hospital settings where already vulnerable patients are exposed to strains of bacteria upon which conventional antibiotics have no effect.
It has been widely suggested that in order to combat this emergence of resistance, research focus needs to shift from developing new antibiotics to novel treatment strategies that target bacterial virulence: an anti-virulence drug that promotes natural clearance through a weakening of their ability to counteract the immune response.
This has given rise to much discussion of the potential success of such an approach: would these drugs be capable of clearing an infection or would they be better suited to infection prevention?
Since these drugs are still in the very early stages of development, we have developed a mathematical model to test their potential efficacy \textit{in silico}.

Our model suggests that, if used in isolation, an anti-virulence drug may not be capable of clearing an infection as it relies too heavily  on the host immune system which was not strong enough to clear the infection itself in the first place (hence the need for treatment).
The immune-related parameters are highly host-specific, though, and if taken to be representative of a healthy individual then the anti-virulence drug could well deal with an infection, suggesting that the drug could be useful for prophylaxis.
However, since antibiotic resistance is mainly a problem in hospital settings, where patients are more vulnerable, this dependence on the immune system is likely to be a limitation of the proposed treatment. 
Nevertheless, regardless of not being able to completely clear the bacterial infection in such immunocompromised patients, we do find some promising results. 
The persisting bacterial load is not only at a much lower level than the model predicted after antibiotic treatment, but importantly the antibiotic-resistant bacteria were eliminated (due to the fitness cost associated with resistance) leaving only antibiotic-susceptible bacteria at the infection site.
These results led us to investigate the potential of combination therapy.
Treatment by both anti-virulence drugs and antibiotics could fully clear an infection if the fitness cost associated with antibiotic-resistance was sufficiently high.
If this cost was lower, combination therapy could still be successful through the imposition of a time delay between treatment types, but the success of this approach was sensitive to the choice of delay between the drug types.

While underpinning the potential of anti-virulence drugs in a theoretical framework, this study also highlights the need to better understand patient- and site-specific parameters pertinent to an infection before optimal treatment strategies can be predicted.
Mathematical modelling is an extremely useful tool for predicting such things, but the results will be dependent on specific parameters within an infection, for example the level of resistance the infecting strain displays to a particular antibiotic, the likely bacterial load at an infection site or the time taken for treatment to reach the infection site.
Thus, further work remains to parameterise the model to given pathogens, infection sites and patients.
Furthermore, the model can be extended to be anti-virulence drug-specific.
For example, anti-adhesive drugs that prevent bacteria binding to host cells should increase the natural clearance rate, rather than promoting the innate immune response, and this can be accounted for in extensions to the model.
Such efforts will greatly enhance the predictive powers of this type of model and contribute to the future design of effective treatment regimen for bacterial infections.

\section{Summary}

Antibiotic resistance in bacteria has long been recognised as a problem for effective treatment of infections, yet it is only more recently that the urgent need for antibiotic alternatives has become widely accepted. 
We have presented here a model of a generalised treatment strategy for changing the target of the drug to promote natural clearance.
Our results support the continuation of research into anti-virulence drugs both in the context of treating infections and in aiding with the elimination of the spread of antibiotic-resistant bacteria from an infection and through hospital settings.

Throughout our analysis of these models we have highlighted some important issues that need to be considered when designing treatment strategies such as the importance of tailoring treatment to specific conditions of the host and how the success of combination therapy (use of more than one drug type) depends highly on the dosing schedule.
Most importantly, however, we have shown conditions under which treatment strategies will be successful and hence, provided a proof of concept for the potential such treatments should have to eliminate the problem of antibiotic resistance.

\textbf{Acknowledgements} RJD gratefully acknowledges the support of the University of Birmingham's System Science for Health initiative. SJ thanks the Medical Research Council for a Biomedical Informatics Fellowship. AMK was funded by a BBSRC New Investigator Grant and an EMBO Long-Term Fellowship.

\bibliographystyle{asm_unsrt}
\bibliography{Bibliography_new}

\end{document}